\documentclass[10pt]{emulateapj}

\def\oiii{{[OIII]}}
\def\nii{{[NII]}}
\def\hb{{H$\beta$}}
\def\ha{{H$\alpha$}}

\begin{document}
\title{\bf The Environment of AGNs in the Sloan Digital Sky Survey\footnotemark[1]} 
\footnotetext[1]{Based on observations obtained with the Sloan Digital Sky Survey}

\author
{Christopher J. Miller,$\!\!$\altaffilmark{2,3} Robert ~C.~Nichol,$\!\!$\altaffilmark{2} Percy L.~Gomez,$\!\!$\altaffilmark{2}
Andrew ~M.~Hopkins,$\!\!$\altaffilmark{4,5} and
Mariangela ~Bernardi~$\!\!$\altaffilmark{2,4}
\\
}

\altaffiltext{2}{Department of Physics, Carnegie Mellon University, 5000 Forbes Avenue, Pittsburgh PA 15217}
\altaffiltext{3}{E-mail: chrism@cmu.edu}
\altaffiltext{4}{Department of Physics and Astronomy, University of Pittsburgh, Pittsburgh PA 14217}
\altaffiltext{5}{Hubble Fellow}

\begin{abstract} 

We present the observed fraction of galaxies with an Active Galactic Nucleus
(AGN) as a function of environment in the Early Data Release of the Sloan
Digital Sky Survey (SDSS). Using 4921 galaxies between $0.05 \le z \le 0.095$,
and brighter than ${\rm M_{r^*}} = -20.0$ (or ${\rm M^{\star}}+1.45$), we find
at least $\sim 20\%$ of these galaxies possess an unambiguous detection of an
AGN, but this fraction could be as high as $\simeq 40\%$ after we model the
ambiguous emission line galaxies in our sample.  We have studied the
environmental dependence of galaxies using the local galaxy density as
determined from the distance to the $10^{th}$ nearest neighbor.  As expected,
we observe that the fraction of star--forming galaxies decreases with density,
while the fraction of passive galaxies (no emission lines) increases with
density. In contrast, the fraction of galaxies with an AGN remains constant
from the cores of galaxy clusters to the rarefied field population.  We
conclude that the presence of an AGN is independent of the disk component
of a galaxy.  We have extensively tested our results and they are robust
against measurement error, definition of an AGN, aperture bias, stellar
absorption, survey geometry and signal--to--noise.  Our
observations are consistent with the hypothesis that a supermassive black hole
resides in the bulge of all massive galaxies and $\simeq 40\%$ of these black
holes are seen as AGNs in our sample. A high fraction of local galaxies with
an AGN suggests that either the mean lifetime of these AGNs is longer than
previously thought ({\it i.e.}, $\ge 10^8$ years), or that the AGN burst more often
than expected; $\sim 40$ times over the redshift range of our sample.

\end{abstract}


\section{Introduction} 

A fundamental goal of observational cosmology is to determine the
spatial distribution of galaxies as a function of galaxy properties,
{\it e.g.}, luminosity, morphology, star formation rate {\it etc.}, as
such observations place constraints on models of galaxy formation and
evolution. In recent years, there has been an explosion in such work
due to improvements in the theoretical modeling of galaxy evolutionary
processes ({\it e.g.}, Kauffmann et al. 1999; Benson et al. 2002)
and the availability of large amounts of high
quality survey data, {\it e.g.}, Sloan Digital Sky Survey (SDSS; York
et al. 2000), 2dF Galaxy Redshift Survey (Colless et
al. 2001), Las Campanas Redshift Survey (Shectman et al. 1996)
and the Century Survey (Wegner et al. 2001).

In this paper, we focus on one aspect of such research, namely the relation
between the fraction$\footnote{Throughout this work, we measure the AGN
fraction as the number of galaxies identified with an AGN over the total
number of galaxies in our sample.}$ of local galaxies that possess an Active
Galactic Nucleus (AGN) and the local environment of those galaxies.  Such
observations can provide new constraints on models of galaxy evolution ({\it
e.g.}, how the AGN may affect other galaxy properties like star--formation),
as well as providing insights into the physical mechanism(s) powering the
AGN. For example, an AGN could be fueled by the same cold gas in the disk
component of galaxies that is driving the star--formation rate (SFR) of those
galaxies. In this case, the existence of an AGN might diminish in dense
regions, analagous to the SFR-density relation (see Lewis et al. 2002; Gomez et
al. 2003, Carter et al. 2001).  Alternatively, several authors have proposed
that galaxy--galaxy collisions fuel the AGN activity by driving gas into the
cores of galaxies and thus onto the black hole (Gunn 1979; Shlosman 1990). A
third possible empirical model proposed by Kormendy \& Gebhardt (2001)
suggests that supermassive black holes are solely a bulge phenomena and were
formed at the same time as the bulge component of galaxies at an much earlier
epoch in the Universe. In this model, the distribution of AGNs would then
trace the distribution of bulges, under the assumption that AGNs are powered
by supermassive black holes (Lynden--Bell 1964).

Therefore, by studying the fraction of galaxies that possess an AGN,
as a function of environment, one can obtain constraints on such
models of AGN activity and formation. Observationally, the fraction of
galaxies with an AGN has increased as the quality and quantity of the
spectral data has improved. In the early work of Dressler, Thompson, \&
Shectman (1985), they found that only a few percent of galaxies
possessed an AGN (see also Huchra \& Burg 1992), and there were
approximately five times as many AGNs in the field compared to the
cores of clusters. In recent years, Carter et al. (2001; CFGKM
hereafter) measured an AGN fraction of 17\%, while Ho, Fillipenko
\& Sargent (1997) find an AGN fraction of $\sim 40\%$ (this difference
depends upon the details of the spectral definition of an
AGN). Furthermore, there is now less evidence for an environmental
dependence on the AGN fraction (see CFGKM).

Herein, we present the {\it AGN--density} relation for galaxies in the
Early Data Release (EDR) of the Sloan Digital Sky Survey (SDSS). This
relation is analogous to the {\it SFR--density} (Hashimoto et
al. 1998, CFGKM) and {\it morphology--density} (Dressler et al. 1980;
Postman \& Geller 1984; Dressler et al. 1997) relations we have
presented elsewhere in Gomez et al. (2003) and Goto et al. (in prep.)
respectively. The work presented herein builds upon earlier
studies of the spectral properties of galaxies as a function of
environment ({\it e.g.}, CFGKM, Tresse et al. 1999, Loveday et
al. 1999). In Section \ref{data}, we discuss the data and techniques
used for identifying an AGN in the SDSS galaxies. In Section
\ref{density}, we present our results on the {\it AGN--density}
relation, while in Section \ref{discussion}, we discuss these results
and their physical implications. We conclude in Section
\ref{conclusions}. Throughout this paper, we used ${\rm H_o}=75 {\rm
km\,s^{-1}\,Mpc^{-1}}$ and $\Omega_{matter}=0.3$ and
$\Omega_{\Lambda}=0.7$.

\section{The Data}
\label{data}
The data are taken from the Early Data Release of the SDSS
(see Stoughton et al. 2002; see also 
Blanton et al. 2003; Fukugita et al. 2003; Gunn et al. 1998; Hogg et al. 2001; Pier et al. 2003;
Smith et al. 2002; Strauss et al. 2002). One of the unique features of the SDSS is
the sophisticated data analysis software pipelines used to reduce the
raw images and spectra into large catalogs of sources and measured
properties. Briefly, for each flux--calibrated SDSS spectrum, a
redshift is determined from both the absorption lines, via
cross-correlation, and the emission lines, via a wavelet--based
peak--finding algorithm (see Frieman et al. in prep).  Once the
redshift is known, the spectroscopic pipeline estimates the continuum
emission at each pixel by using the median value seen in a sliding box
of 100 pixels centered on that pixel.  Emission and absorption lines
are measured by fitting a Gaussian, above the continuum, at the
redshifted rest--wavelength of the lines. In order to avoid line
blending, the SDSS pipeline fits multiple Gaussians in such cases as
\nii~ and \ha\,.  Thus, the spectroscopic pipeline provides a high
quality estimate of the equivalent width (EW), continuum,
rest--wavelength, identification, goodness--of--fit ($\chi^2$), height
and sigma (and the associated errors on these quantities) for all the
major emission and absorption lines in galaxy spectra. This
extensive emission and absorption line database, when combined with
color and morphological information from the SDSS photometric survey,
is ideal for studying the properties of low redshift galaxies.

The galaxy sample we use is similar to the one described in Gomez et
al. (2003).  Briefly, we define a pseudo volume-limited sample by restricting
the analysis to galaxies in the redshift range $0.05\le~z\le 0.095$, and to
galaxies more luminous than a (k--corrected; Blanton et al. 2002) $M(r^*) =
-20$ which is $M^{\star}(r^*)-1.45$ assuming $M^{\star}{r^*}~=~-20.8+5\log{h}$
from Blanton et al. (2001).  The lower redshift limit is to minimize aperture
bias (see Gomez et al. 2003 and Section \ref{aperture}) due to large nearby
galaxies, while the upper redshift limit corresponds to where luminosity limit
equals the apparent magnitude limit of the SDSS ($r^*=17.77$; Strauss et
al. 2002).

\subsection{Survey Edges}

The SDSS EDR covers a contiguous area of $\simeq 400$ square degrees with a
complicated geometry. Much of the area is contained within two long stripes
that are $2.5$ degrees wide in declination, centered on the celestial
equator. Thus, many of the galaxies in the EDR are close to a survey
boundary. This geometry requires that careful attention be paid to any measure
of local galaxy density measurements.  We have chosen to use the distance to
the 10$^{th}$ nearest neighbor (see Section \ref{density}) to infer local
densities.

\begin{figure}
\plotone{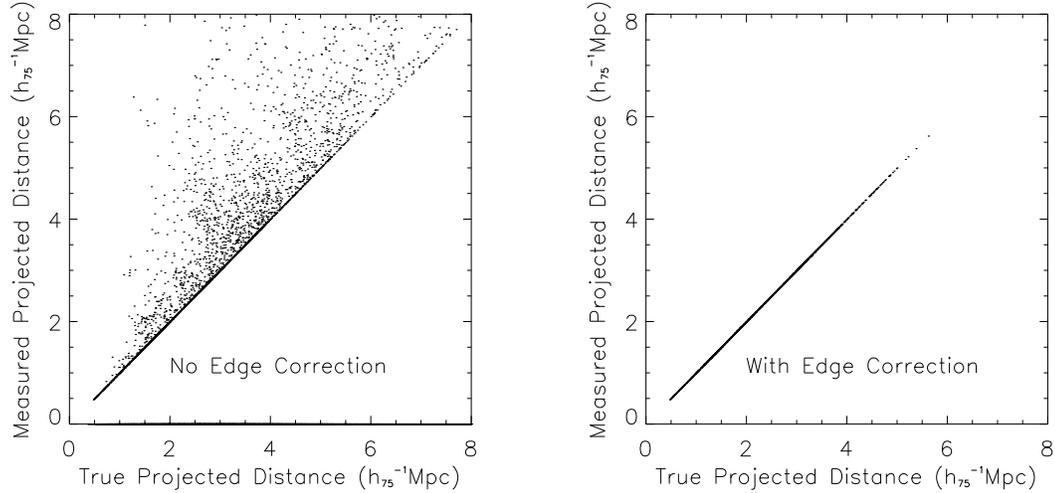}
\caption[]{We present here our analysis of the effect of edge correction on
our data set using the mock galaxy catalogs produced by Cole et
al. (1999). (Left) We compare here the true distance to the $10^{th}$ nearest
neighbor, taken from the mock galaxy catalog without edges, to the measured
$10^{th}$ nearest neighbor, taken from a sample of mock galaxies with the same
survey geometry as the SDSS EDR. (Right) The same data as shown in the
left-hand panel, but now with our edge correction applied as discussed in the
text, {\it i.e.}, we remove all galaxies with an edge closer than the
$10^{th}$ nearest neighbor. As one can see, our edge correction removes the
biased galaxies in the left-hand panel, but does reduce the dynamic range of
densities probed by our analysis, {\it i.e.}, we are restricted to galaxies
with a $10^{th}$ nearest neighbor distance of less than $\sim 6{\rm
h_{75}^{-1}}\, {\rm Mpc}$, or a range of densities from $\sim 10{\rm
h_{75}^2}\, {\rm Mpc^{-2}}$ to $\sim 0.1{\rm h_{75}^2}\, {\rm Mpc^{-2}}$. }
\label{fig::density_bias}
\end{figure}

To avoid a biased density measure, we have excluded galaxies that have a
survey edge closer than the distance to the 10$^{th}$ nearest neighbor.  We
have tested this method of ``edge correction'' using the mock galaxy catalogs
of Cole et al. (1999). We used these data to create a mock galaxy catalog with
the same geometry as the SDSS EDR, and then measure the projected distance to
the 10$^{th}$ nearest neighbor in the full mock ({\it i.e.}, without edges)
and within the catalog with the EDR geometry. Our results are shown in Figure
\ref{fig::density_bias}, which shows that if we ignore edges, the measured
distance to the 10$^{th}$ nearest neighbor is biased towards higher values
compared to the true 10$^{th}$ nearest neighbor distance. However, as
expected, once we remove all galaxies which have an edge closer than the
10$^{th}$ nearest neighbor, we then recover the true distances, or local
densities.  Figure \ref{fig::density_bias} shows the importance in correctly
accounting for edges in surveys with complicated volumes, such as the SDSS
EDR.

\begin{figure}
\plotone{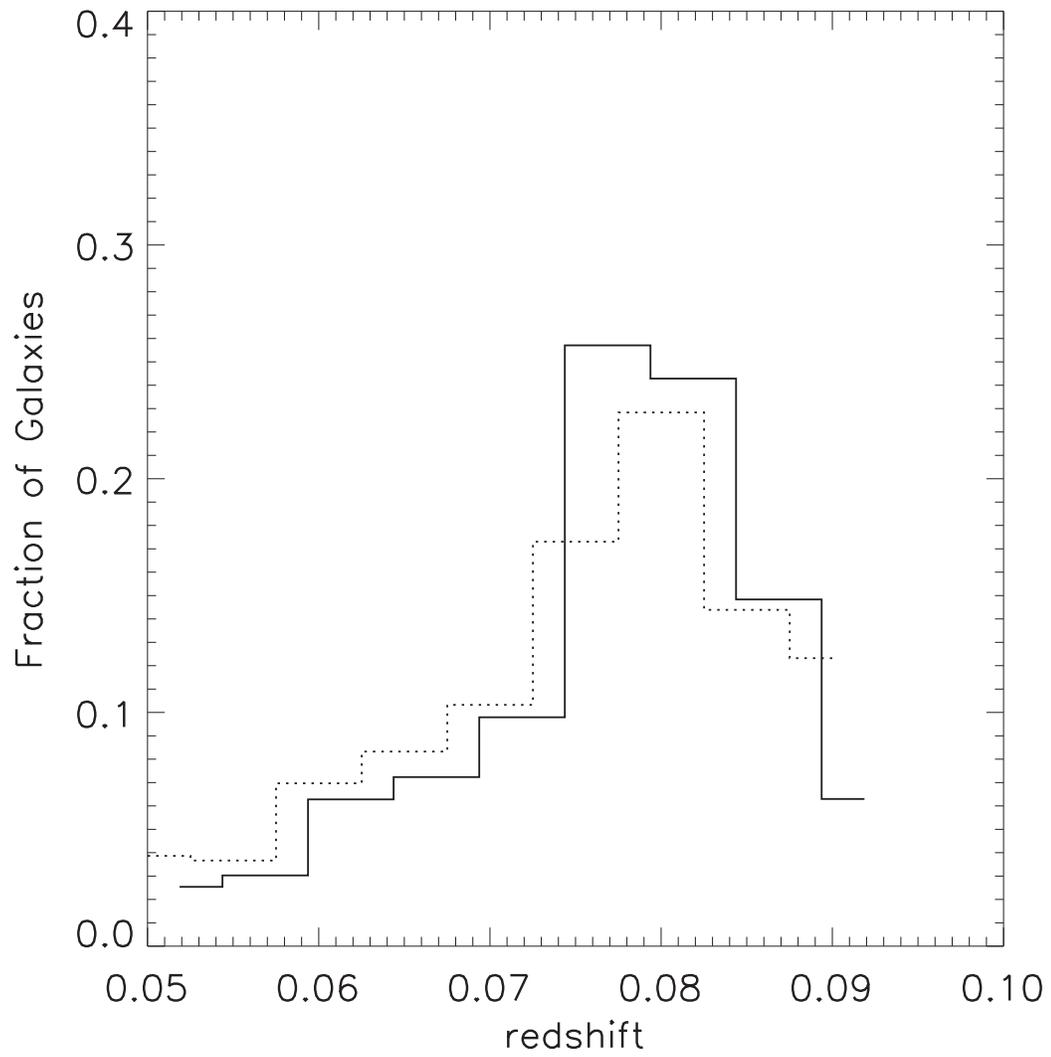}
\caption[]{We show here the normalised redshift histograms for the whole SDSS
EDR catalog (within our absolute magnitude limits; dotted line) and for our
sample after edge-corrections (solid-line).}
\label{fig::zhisto}
\end{figure}

We note that while the edge correction produces an unbiased estimate of the
density (as shown in Figure \ref{fig::density_bias}), the {\it distribution}
of densities for our sample is different than the whole EDR survey. In Figure
\ref{fig::zhisto}, we plot the redshift histogram of our sample before and
after the edge correction is made. The edge correction results in a smaller
survey area (than the whole EDR) at the lowest redshift, as the volume sampled
at $z\simeq0.05$ is small compared to the typical distance to the 10$^{th}$
nearest galaxy neighbor).  Additionally, since galaxies in dense environment
have closer neighbors than galaxies in voids, we preferentially sample dense
regions more often than sparse regions; we note however that the
signal-to-noise of the measured densities is independent of this effect as we
alway use 10 galaxies to define the local galaxy density. 

Our results are robust against this bias in the {\it distribution} of
densities for two reasons. First, we only study herein the {\it fraction of
galaxies} as a function of the various galaxy types ({\it i.e.}, AGN,
star-forming and passive) and have binned the galaxies in density such that
each bin has $\simeq 250$ galaxies per bin (this changes slightly for the
analyses as a function of redshift). Therefore, higher density regions are
more finely binned compared to lower density regions.  Secondly, we have
divided our sample into three nearly equal volume subsets, as a function of
redshift, and show that our conclusions are independent of redshift (see
Figure \ref{fig::bias}). This suggests that the bias seen in Figure
\ref{fig::zhisto} is unimportant. In summary, our tests show that we possess
an unbiased estimate of the local (projected) galaxy density over two orders
of magnitude in the density, {\it i.e.}, from $\sim 10{\rm h_{75}^2}\, {\rm
Mpc^{-2}}$ to $\sim 0.1{\rm h_{75}^2}\, {\rm Mpc^{-2}}$ (see Figure
\ref{fig::density_bias}).  Our final, edge-corrected sample, contains 4921
galaxies with a mean spectral signal-to-noise ratio$\footnote{The
signal--to--noise is measured as a mean per pixel in the SDSS photometric
r-band (Stoughton et al. 2002)}$ for their spectra of 24. The wavelength
coverage is 3900\AA~ to 9100\AA, with a spectral-resolution of 1800.

We note here that there are 27 sources spectrally classified by the SDSS
software pipeline as quasars within the sample limits defined above.  These 27
low redshift QSOs (which are actually Seyfert Type I AGNs) possess extremely
broad lines making it difficult to obtain line flux-ratio measurements
(especially for \nii~ and \ha~ which lie only 20\AA~ apart).  We note that a
Kolmogorov-Smirnov test indicates no difference between the environments of
these low-z QSOs and of our AGN sample defined in Section
\ref{identify}. Therefore, as there are so few of these objects (0.5\% of our
total sample), and they exist in the same environments as our full AGN sample
(of nearly 1000 galaxies), we ignore these QSOs unless otherwise noted.

\subsection{Identifying AGNs: Part I}
\label{identify}

To classify AGNs, we use the traditional log(\oiii/\hb) versus log(\nii/\ha)
flux-ratio diagnostic diagrams as discussed by Baldwin, Philips \& Telervich
(1981), Villieux \& Osterbrock (1987) and Kewley et al. (2001). The flux of
these emission lines are determined using the fitted heights and sigmas of the
lines as measured by the SDSS data analysis pipeline. We initially restrict
our analysis to lines with a measured flux over flux error of $>2$, {\it
i.e.}, $>2 \sigma$ detections. In Figure \ref{fig::diagnostic} (left), we show
the model (solid line) used to separate star--forming galaxies from AGNs.
Galaxies which lie above, and to the right of, this model are classified as
AGNs, while galaxies beneath, and to the left, are classified as star--forming
galaxies.




For the 4921 galaxies in our pseudo volume-limited sample, 3647 have
at least one $>2\sigma$ emission line detection, while 1293 galaxies
have all of the four lines required in Figure \ref{fig::diagnostic}
for an unambiguous classification ({\it e.g.}, the \nii, \oiii, \ha~
and \hb~ emission lines). Most of the galaxies with these four emission
lines (82\%) are classified as star-forming galaxies. These
star-forming galaxies have a median \ha~ equivalent width (EW) of
26\AA. The remaining 235 galaxies are classified as AGNs with a
median \ha~ EW of 13\AA.  We call these the ``4--line AGNs'' for the
rest of the paper. We have made no attempt to classify these AGNs into
sub--populations like LINERs, Seyferts I or II.  We simply note that
these 235 AGNs do cover a broad range of AGN types. Likewise, we have
made no attempt to study the individual properties of the AGNs.

We note here that it is not necessary to possess significant detections of all
four of these lines (\nii, \oiii, \ha~ and \hb) to unambiguously detect an AGN
in a galaxy; although one does need all four lines to unambiguously classify
star-forming galaxies.  We illustrate this fact in Figure \ref{fig::separ},
which shows that galaxies with either a high \nii/\ha~, or a high \oiii/\hb~,
line ratio must be an AGN (see CFGKM), regardless of the other line ratio.
[We note that we find no AGNs using solely the  \oiii/\hb~ line ratio].
Star--forming galaxies only inhabit the bottom left--hand corner of Figure
\ref{fig::diagnostic}.  For example, if a galaxy has \nii~ and \ha~ lines
detected at $>2\sigma$ confidence, but no significant detection of \oiii~ or
\hb, we can still classify it as an AGN if log(\nii/\ha)$>-0.2$.  Carter et
al. (2001) used a similar technique to extract additional AGNs in their data.
To differentiate these AGNs from the 4--line AGNs discussed above, we call
these AGNs ``2--line AGNs'', as they are classified using only two
lines. These ``2--line AGNs'' have much lower line strengths than the
``4--line AGNs'', as they possess a median EW of H$\alpha)\simeq 3$. However,
there is no difference in the signal-to-noise ratios of the ``2--line''
AGN spectra compared to ``4--line'' AGN spectra. In Figure
\ref{fig::llagn}, we show the \nii/\ha~ flux-ratios for these 2--line AGNs.



Using log(\nii/\ha)$>-0.2$, we have increased our AGN sample from 235
(4--line AGNs) to 936 galaxies. Thus, the majority of our AGNs are
classified using only the \nii~ and \ha~ lines, which has two important
advantages compared to the 4--line method.  First, the classification
is unambiguous, since we have chosen line ratios high enough to
exclude nearly all star--forming galaxies in Figure 1. Second, \nii~
and \ha~ are close together in wavelength so we have no concerns about
the amount of internal dust extinction when classifying these galaxies.

There remains a large number of
galaxies that have at least one $>2\sigma$ detection of an emission line,
but cannot be classified using either the 2--line or 4--line method
discussed above. We call these galaxies ``Emission Line but
Unclassified'' (ELU) galaxies. We discuss these further in Section
\ref{ELU}.  Any galaxy with no $>2\sigma$ detected emission lines is
classified as ``passive''.  In Table 1, we present the fraction of
galaxies as a function of these different classifications.

\subsection{Identifying AGNs: Part II}

To study the effect of measurement errors
(see CFGKM), we apply a different classification technique to our sample,
which utilizes the observed error on the line ratios, and calculates
the probability that each galaxy is either star--forming or contains
an AGN. The method uses the False Discovery Rate (FDR) described in
Miller et al. (2001) and has the advantage of controlling the amount
of contamination in each class of galaxies, which is the
scientifically meaningful quantity. For our FDR technique, we drop the
requirement that all lines must be detected at the $>2\sigma$
confidence level, and instead use all lines that have a positive value
({\it i.e.}, emission) regardless of their measurement error.

To use FDR, we must begin by assuming a null hypothesis and test each
galaxy against that null hypothesis. Our first null hypothesis is to
assume that any galaxy below the model line shown in Figure
\ref{fig::diagnostic} is a star--forming galaxy, {\it i.e.}, it has
a probability of 1 of being star--forming. Then, for all galaxies
above the line in our sample, we assign a probability of being
star--forming assuming the errors on the line ratios are Gaussian.
Once we have a probability for each galaxy, we use FDR to reject
galaxies from this null hypothesis ({\it i.e.}, they are AGNs) at a
confidence level of 90\%.  Therefore, we obtain a sample of AGNs that
is guaranteed to have less than 10\% contamination from star-forming
galaxies.  We then alter our null hypothesis, so that all galaxies
above the model line in Figure \ref{fig::diagnostic} are classified as
AGNs, and re--calculate the probability for each galaxy below the
line. Once again we use FDR to control the false detection rate and
obtain a sample of star--forming galaxies that are rejected based on
this new null hypothesis. Again, this guarantees fewer than 10\%
contamination by AGNs in our sample of star--forming galaxies. This
procedure produces a robust sample of star--forming and AGN
classifications, as well as galaxies that are unclassified because
they were rejected against both null hypotheses.

In Figure \ref{fig::diagnostic} (right) we plot the flux ratio
diagnostic diagram constructed using the FDR method described
above. In this figure, all galaxies with line ratios close to the
model separating AGNs and star--forming galaxies are unclassified. The
FDR classification scheme (which properly treats the errors on the
line-fluxes and ratios) picks up more AGN with weak lines ({\it i.e.}
some less than 2$\sigma$) at the expense of missing ``transition
galaxies'' which have evidence of both star-formation and an AGN.
Overall, the FDR method is more conservative than the the $>2\sigma$
line method described in Section \ref{identify}.  However, we still
find a similar fraction of AGNs and star--forming galaxies as we did
when simply requiring all lines were detected at the $>2\sigma$
confidence. For example, the total number of 4--line and 2--line AGNs
we calculate using the FDR method is similar to that found using the
$>2\sigma$ method; 14\% for FDR compared to 19\% in Table 1.

\subsection{Low-Luminosity AGNs}
\label{llagn}

The SDSS spectroscopic analysis pipeline (see Section \ref{data}) determines
the location and flux of an emission line via a Gaussian fit to the spectral
data. This fit does not account for the possibility of stellar absorption
which would lie underneath any emission line.  Such absorption could lead to a
systematic uncertainty in the measured flux of the hydrogen emission lines in
the SDSS spectra (see Gomez et al. 2003; Goto et al. 2003). To overcome such
problems, several authors (Ho, Fillipenko, Sargent 1997; Coziol et al. 1998)
have advocated using template spectra to estimate the amount of stellar
absorption underneath the hydrogen lines. Such a technique involves fitting
the continuum flux of each galaxy spectrum with these templates and finding
the best fit. Then, the emission lines are measured taking into account the
expected amount of stellar absorption as seen in the best--fit template
spectrum.  This technique has the advantage of being more sensitive to lower
luminosity AGNs (LLAGNs), as one is able to identify and measure the weaker
emission lines. This, in part, explains why Ho, Fillipenko, Sargent (1997) and
Coziol et al. (1998) find more AGNs than previous studies who did not use
template fitting techniques.

We examine here the issue of low luminosity AGNs and test our ability of
finding such sources using our AGN identification techniques as discussed in
Section \ref{identify}.  As a basis for our tests, we compare our analysis
with the sample of 82 Hickson group galaxies studied by Coziol et al. (1998),
who used the template fitting technique to study the LLAGN population in
groups of galaxies. As we see no dependence on the AGN fraction with
environment (see Section \ref{density}), the environment of the Coziol et
al. sample is irrelevant and does not bias our tests.  In Figure
\ref{fig::llagn}, we present the observed rest-frame equivalent width of \nii~
for galaxies in our sample as a function of their observed \nii/\ha~ line
ratio. Only galaxies with $>2\sigma$ detections of both the \nii~ and \ha~
lines are shown here. Figure \ref{fig::llagn} is therefore similar to Figure 5
in Coziol et al. (1998), who use such a diagram as a diagnostic to indicate
the separation between star--forming galaxies, AGNs and LLAGNs.

We find a significant population of our 2--line AGNs are LLAGNs, as
they possess weak \nii~ and \ha~ emission lines, {\it e.g.}, only a few
\AA~ of EW in these detected lines. This is because of the high
resolution of the SDSS spectra (2.3\AA~ per resolution element), which
helps de--blend the \nii~ doublet and \ha~ emission lines, and the high
mean signal--to--noise ratio (S/N$=23.9$) of our LLAGN spectra. Therefore,
we have the sensitivity to detect any strong \oiii~ and \hb~ emission in
these spectra if it was present. In summary, most of our 2--line AGN
detections are LLAGNs, as by definition they possess weaker \nii~ and \ha~
emission lines (but with a large line ratio), and no detectable \oiii~
and \hb~ emission. 

We note here that we find no correlation between our
\ha~ fluxes and the signal-to-noise of the spectra. In other words,
it is not the case that all of our low \ha~ flux galaxies have low
S/N spectra (and thus large errors). 

\subsection{Effects of Stellar Absorption}
Our ability to detect LLAGNs (discussed above) and our
agreement with
Coziol et al. (1998) indicates that our method
can accurately classify AGNs without the need for template fitting.
To expand on this, we have performed two tests quantify the effect
of stellar absorption on our classifications.

First, we have visually inspected the profiles of the \nii~ and \ha~ emission
lines in the spectra of the 50 AGNs with the smallest \ha~ equivalent widths
that are close to the model line in Figure \ref{fig::llagn}.  Using the
absorption in \hb~ as an upper limit to the total possible stellar absorption
in the galaxy, we conclude that $\sim 5$ of these 50 galaxies could change
classification from AGN to star-forming.  These inspections demonstrate that
only $10$\% of these lowest luminosity AGNs are affected by stellar
absorption. Second, we add 0.7\AA~ of stellar absorption to all the hydrogen
lines in our spectra and re-classified them. This mean amount of stellar
absorption has been estimated by Hopkins et al. (in prep) to obtain consistent
measurements of the SFR of SDSS galaxies using H$\alpha$, [OII], $u'$-band
luminosity, with the IRAS infrared luminosities and radio luminosities from
the FIRST survey.  After applying this absorption correction, our total
fraction of AGNs changes from 19\% (in Table 1) to 17\%.  We conclude from
both of these tests that our final sample of ``2--line'' and ``4--line'' AGNs
is robust against stellar absorption corrections. Since we are not attempting
to measure precise line fluxes in this work, we ignore stellar absorption
throughout the rest of this work.

\subsection{Aperture Effects}
\label{aperture}

We investigate here aperture effects due to the limited radius of the SDSS
spectroscopic fibers (only 1.5 arcseconds in radius), using the SDSS
photometric data to measure the fraction of light contained within the fiber
(1.5 arcseconds), compared to the total light of a galaxy, as defined by the
Petrosian 90\% radius (see Stoughton et al. 2001 for more details). We begin
by noting that this fraction is approximately the same for all galaxy types
discussed herein, {\it i.e.}, we find the fraction of the total light within
the fiber to be 25\% for star-forming galaxies, 23\% for AGN, and 27\% for
passive galaxies. Thus, over our redshift range, our spectral identifications
are not biased towards any specific galaxy type.

In Figure \ref{fig::han2_agn_v_radius}, we plot the measured [NII]/H$\alpha$
line ratio, as a function of the fraction of the total light within the SDSS
fiber, for galaxies classified as either an AGN or a star--forming galaxy.  We
see no evidence for any aperture bias for either AGNs or star-forming samples
(nor as a function of luminosity). In Figure \ref{fig::han2_vz}, we plot the
measured [NII]/H$\alpha$ line ratio as a function of redshift and again, see
no evidence for aperture bias with redshift.



\subsection{Completeness} 

The main SDSS galaxy sample is designed to be near complete to our chosen
magnitude limit (see Strauss et al. 2002) and therefore, we expect our pseudo
volume--limited sample of galaxies to be nearly-complete in both redshift and
absolute magnitude within the limits specified in Section \ref{data}. The use
of an absolute magnitude limited survey is crucial for future comparisons to
theoretical work (via simulations or semi-analytic techniques). We stress that
our AGN fraction as determined using all unambiguous classifications (19\% as
seen in Table 1) is relatively robust against the chosen magnitude limit for
our sample. For instance, the AGN fraction increases only slightly, from 19\%
to 22\%, over the magnitude range $-19 \le M_{r} \le -22$ respectively. We
have difficulty probing magnitudes which are brighter (too few galaxies) or
dimmer (the sample becomes very incomplete) than these limits.

It is also worth pointing out that the SDSS has a firm lower limit on the
signal-to-noise ratio of any observed spectrum, which alleviates any concerns
regarding the homogeneity of identifying emission lines in these spectra.  To
test this, we measure the mean signal--to--noise ratio for each of the
different classification of galaxies in our sample and found that the passive
population (no emission lines greater than 2$\sigma$) had the highest mean
signal--to--noise of all the sub--samples, {\it i.e.}, a mean S/N of 27.7
compared to a mean S/N of 23.8 for the AGNs and a mean S/N of 18.5 for the
star--forming populations.  Therefore, these passive galaxies are truly
passive as they possess the required signal--to--noise to detect the expected
emission lines seen in AGNs, LLAGNs and star-forming galaxies.  The ELU
galaxies ($34$\% of all galaxies in our sample) have a mean signal--to--noise
ratio of 24.8. Therefore, in terms of our galaxy classifications, our total
galaxy sample is 66\% complete (in classifications), with the remaining third
of galaxies being classified as ELU galaxies. We will discuss these ELU
galaxies in Section \ref{ELU}.

\subsection{Data Summary}

For two--thirds of our sample, we have a robust classification of an AGN,
passive and star--forming galaxy. For example, the two statistical methods
used to identify AGNs and star--forming galaxies produce similar
fractions. The first method, which is common in the literature, is less
conservative than our FDR method, and produces a slightly larger sample of
AGNs. However, the differences are marginal, indicating that our galaxy
classifications are robust to the method used and the errors on the line
measurements. The AGN classifications are thus robust against the exact
classification of the transition objects. Therefore, for the remainder of the
paper, we use the $>2\sigma$ line method to classify our galaxies to remain
consistent with the previous literature.

We have compared our AGN classification to those using template-fitting
techniques ({\it e.g.}, Coziol et al. 1998) and find excellent agreement (see
Figure \ref{fig::llagn}).  This is due to our ``2--line'' method of
classifying AGNs, which only requires the stronger \ha~ and \nii~
lines. Finally, we have shown that stellar absorption does not strongly affect
our classifications and that our classifications are free of any aperture
bias.  The remaining third of our sample is unclassified and could be affected
by measurement error, stellar absorption and classification technique. We make
no attempt to individually classify these galaxies, but instead statistically
model their distributions (see Section \ref{ELU}).

\section{AGN Fractions versus Density}
\label{density}

In Figure \ref{fig::agnfrc_dens}, we plot the fraction of galaxies in
each of our classifications (star--forming, passive, AGN, ELU) as a function
of the projected local galaxy density, which is identical to that used, and
described, in Gomez et al. (2003).  Briefly, this projected local galaxy
density is determined from the distance to the $10^{th}$ nearest projected
neighbor (within our pseudo volume--limited sample) within a redshift shell of
$\pm1000{\rm km\,s^{-1}}$ centered on the central galaxy. We then convert this
surface density (within the redshift shell) into a local projected galaxy
density. This approach is adaptive as the width of the smoothing kernel changes as a function of the local density and has the advantage, over a
fixed width kernel density estimator, of having a constant signal--to--noise
for all environments. We show in Figure \ref{fig::bias} that our density
estimates are robust against choices of the size of the redshift shell and
independent of redshift.



Figure \ref{fig::agnfrc_dens} shows the expected decrease in the fraction of
star--forming galaxies in dense regions and the increase in the fraction of
passive galaxies in dense environments.  These relations are analagous to
SFR--density and morphology--density relations of Gomez et al. (2003) and
Dressler et al. (1997) respectively. Figure \ref{fig::agnfrc_dens} confirms an
earlier analysis by CFGKM who used an entirely different dataset and
data-reduction procedure. As in CFGKM, we find that the AGN fraction is
constant over the two orders of magnitude in local galaxy density probed by
this work, {\it i.e.}, spanning a density range that includes the cores of
rich clusters to the rarefied field population. The small decrease in the AGN
fraction in the densest regions is not statistically significant; a line with
zero slope is a good fit to the fraction versus density. 

In addition, we have
performed chi-square tests on the fraction distributions in Figure \ref{fig::agnfrc_dens}
(Left). In this analyis, we measured the chi-square test statistic using
the mean of the fraction per galaxy type as the null hypothesis.
This test then tells us whether we can rule
out a constant fraction (defined by the mean) as a function of density for each
galaxy type. For AGN, we can only rule out a constant AGN fraction at
the 46\% level (i.e., a constant AGN fraction cannot be ruled out for this data).
On the other hand, we can rule out a constant star-forming fraction at the $>$99.9999\%~
level, and a constant fraction of passive galaxies is ruled out at the 99.999\%~ level.
We have looked at this statistic using only 2--line or 4--line AGNs as defined above
and also find that the a mean constant fraction cannot be ruled out for the either of
these subsets of AGNs.
We can restate this result in that the dip in the inner-most bin of the AGN fraction
seen in Figure \ref{fig::agnfrc_dens} (Left) is not significant. We note that there
are a significant number of ELUs in this bin, which must be classified as either
star-forming or AGN.
In the next section,
we address how the ELU galaxies might contribute to the AGN and star--forming
fractions.

\subsection{Identifying and Modeling the ELU Galaxies}
\label{ELU}

Unfortunately, in Figure \ref{fig::agnfrc_dens}, 34\% of our galaxy sample
remains unclassified as either an AGN or star--forming galaxy; these are the
ELU galaxies discussed in Section \ref{identify} and their fraction as a
function of density is shown in Figure \ref{fig::agnfrc_dens}. These ELU
galaxies cannot be passive as, by definition, they contain at least one
significantly detected emission line ($>2\sigma$). In this section, we first
try and classify as many of these ELU galaxies as possible using the available
spectroscopic information. Then, for the remaining ELU galaxies, we
statistically model their contribution to the AGN and star--forming fractions
seen in Figure \ref{fig::agnfrc_dens}, using only our current knowledge of the
environmental dependence of star--forming galaxies.

\subsubsection{Upper Limits on Missing Emission Lines}
We discuss here a technique for classifying ELU galaxies based on the
combination of detected ($>2\sigma$) emission lines and upper limits on the
non--detected emission lines ($<2\sigma$). For example, most of the ELU
galaxies in fact have $>2\sigma$ detections of both the \ha~ and \nii~
emission lines, but are missing detections of the weaker \hb~ and \oiii~
lines.  These galaxies were not classified as AGNs, using the 2--line method
discussed in Section \ref{identify}, as their observed \nii/\ha~ line ratio
was below the threshold required for them to be unambiguous 2--line AGN
detections (see Section \ref{identify} where we defined log(\nii/\ha)$>-0.2$
as the threshold).  However, for these ELU galaxies, we can make a reasonable
estimate of their classification using the observed upper limits on the fluxes
of the \oiii~ and \hb~ emission lines. The flux upper limits on these missing
lines is taken to be twice the measured flux error on that line, as this is
our criteria for considering a line as detected, {\it i.e.}, $>2\sigma$.

\begin{figure}
\plottwo{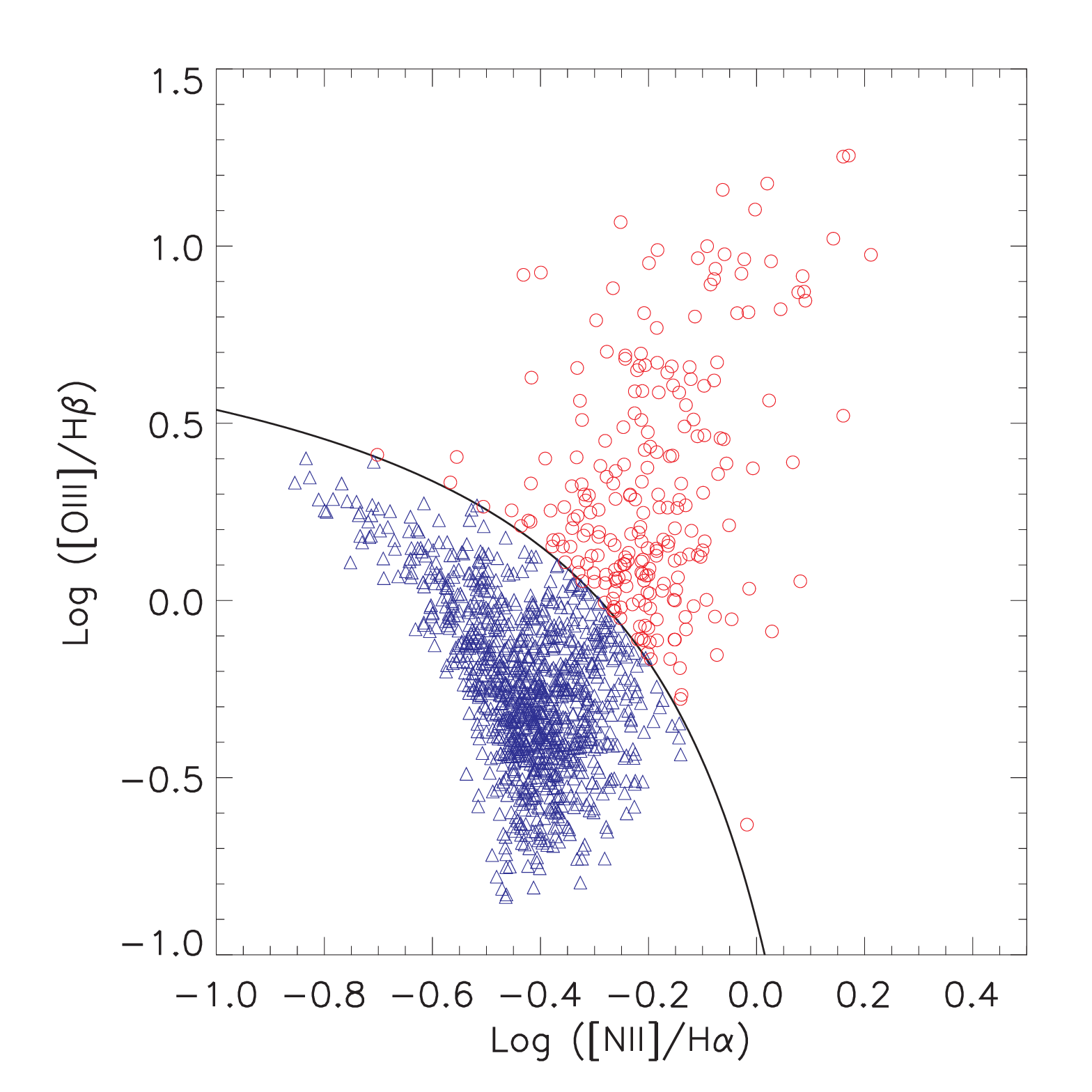}{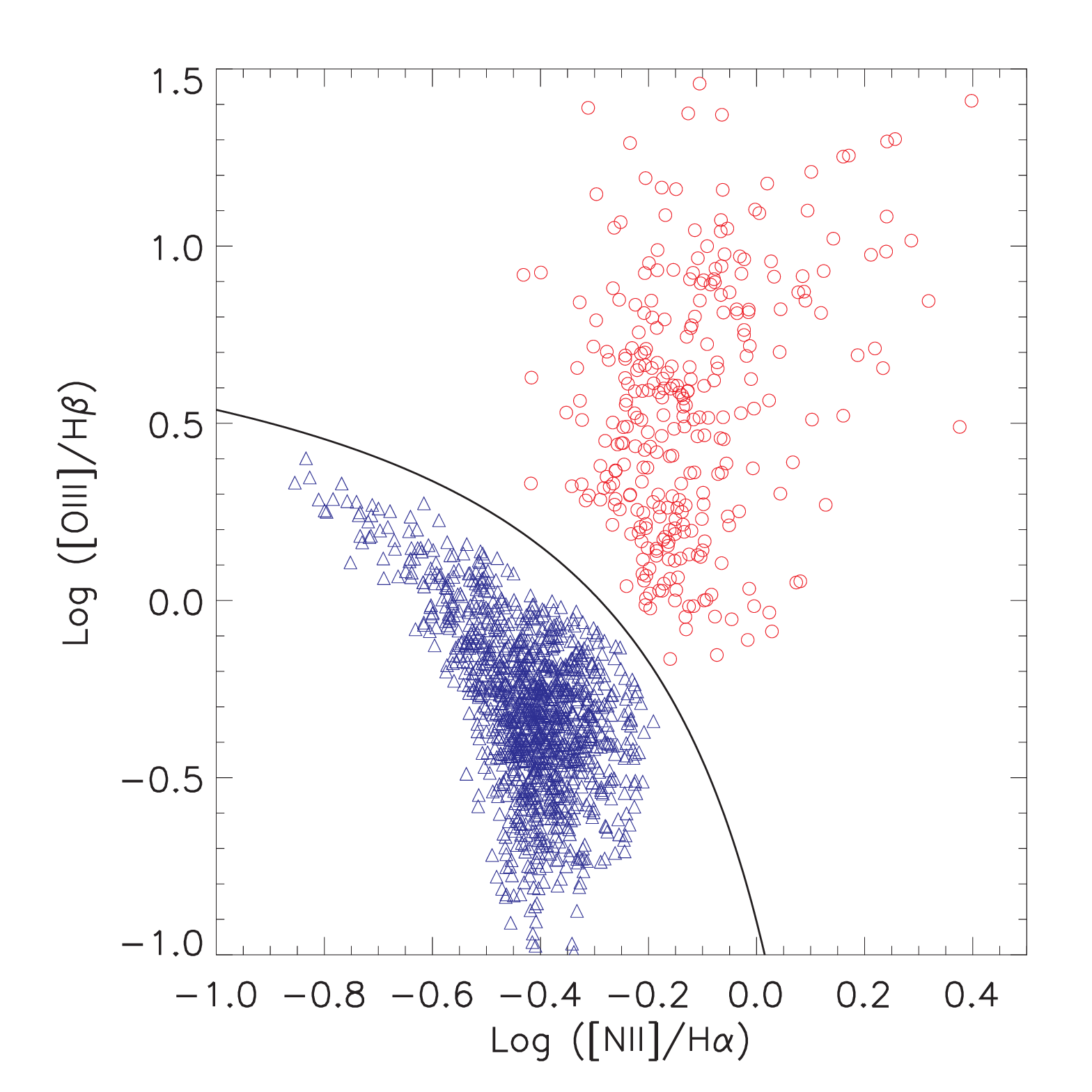}
\caption[]{The line diagnostic diagram traditionally used to
differentiate AGNs (red circles) from star--forming galaxies (blue
triangles).  The solid line is the one--sigma lower limit of the
model from Kewley et al. (2001). (Left) The AGNs in this figure are called
``4--line AGNs'' in the text, as they require all four emission lines
to be present for an
unambiguous classification. (Right) We plot our classifications using
the FDR method described in the text. The FDR method is more
conservative and can not classify the ``transition objects'' ({\it
i.e.}, those galaxies that lie near the model and have a signature of
both star--formation and an AGN). However, the two statistical methods
used to classify AGNs and star--forming galaxies produce similar
fractions.}
\label{fig::diagnostic}
\end{figure}

\begin{figure}
\plotone{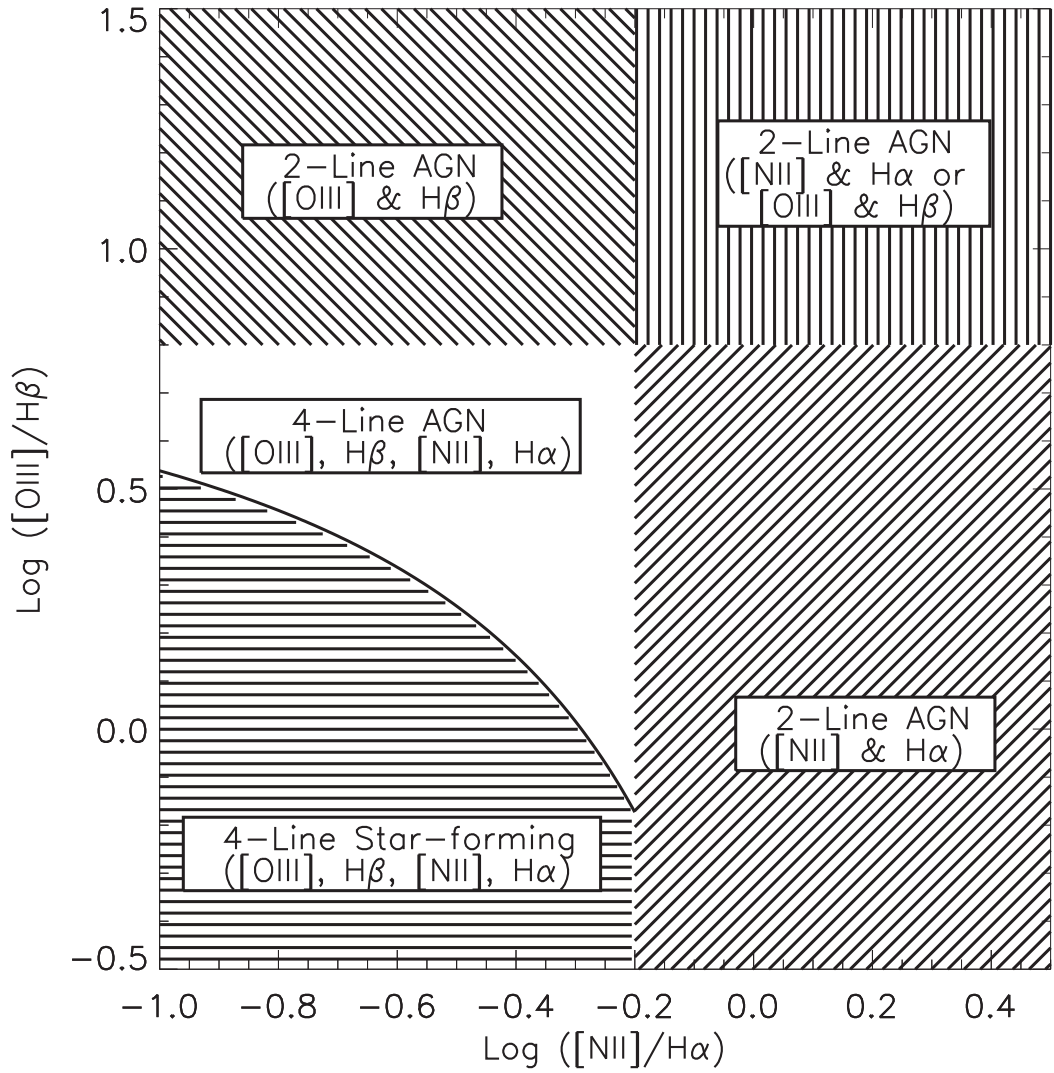}
\caption[]{This schematic shows how we identify AGNs.  The solid curve
is the same model used in Figure \ref{fig::diagnostic}.  }
\label{fig::separ}
\end{figure}

\begin{figure}
\plotone{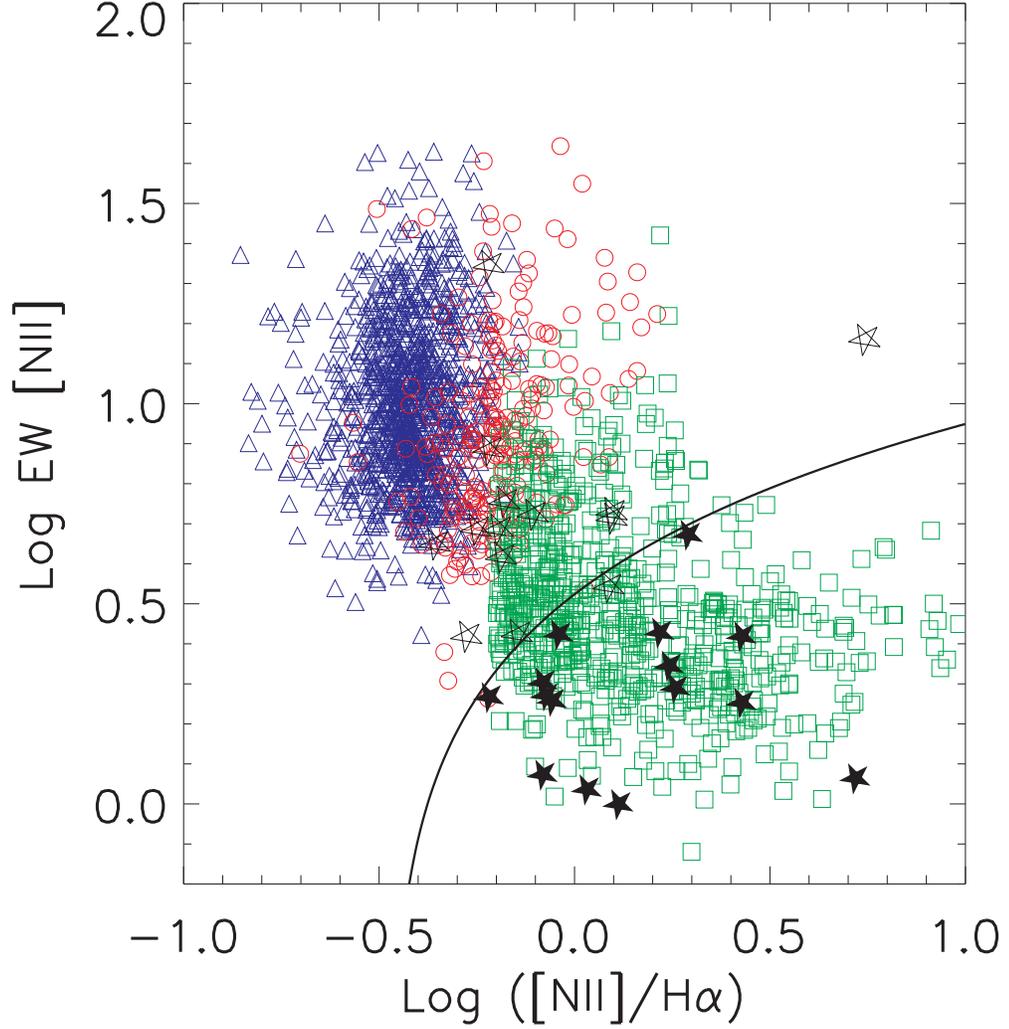}
\caption[]{\small We plot the equivalent width (EW) of the \nii~
emission line versus the line ratio \nii/\ha~ for galaxies in our
sample with $>2\sigma$ detections of these emission lines.  The blue
triangles are unambiguous star-forming galaxies (4-line method),
while the red circles are our 4-line AGN identifications. The green
squares are our 2--line AGNs.  Many of our 2-line AGNs are normal,
{\it i.e.}, they overlap with the 4-line AGNs, while others are low
luminosity AGNs (LLAGNs), delineated using the solid curve. We compare
our data to that of the AGNs identified by Coziol et al. (1998) (shown
as stars). The solid stars were identified as LLAGNs by Coziol
et al. using their template fitting technique, while the open
stars are normal AGNs.  We find a similar fraction of LLAGNs to
normal AGNs in our data as found for Coziol et al. data. We note our data are
incomplete for low luminosity star-forming and 4--line AGNs (in the
region of log(\nii/\ha)$<-0.2$ and log(EW(\ha))$< 0.5$) because we
require a significant detection of all \nii, \oiii, \ha~ and \hb~ emission lines. }
\label{fig::llagn}
\end{figure}

\begin{figure}
\plotone{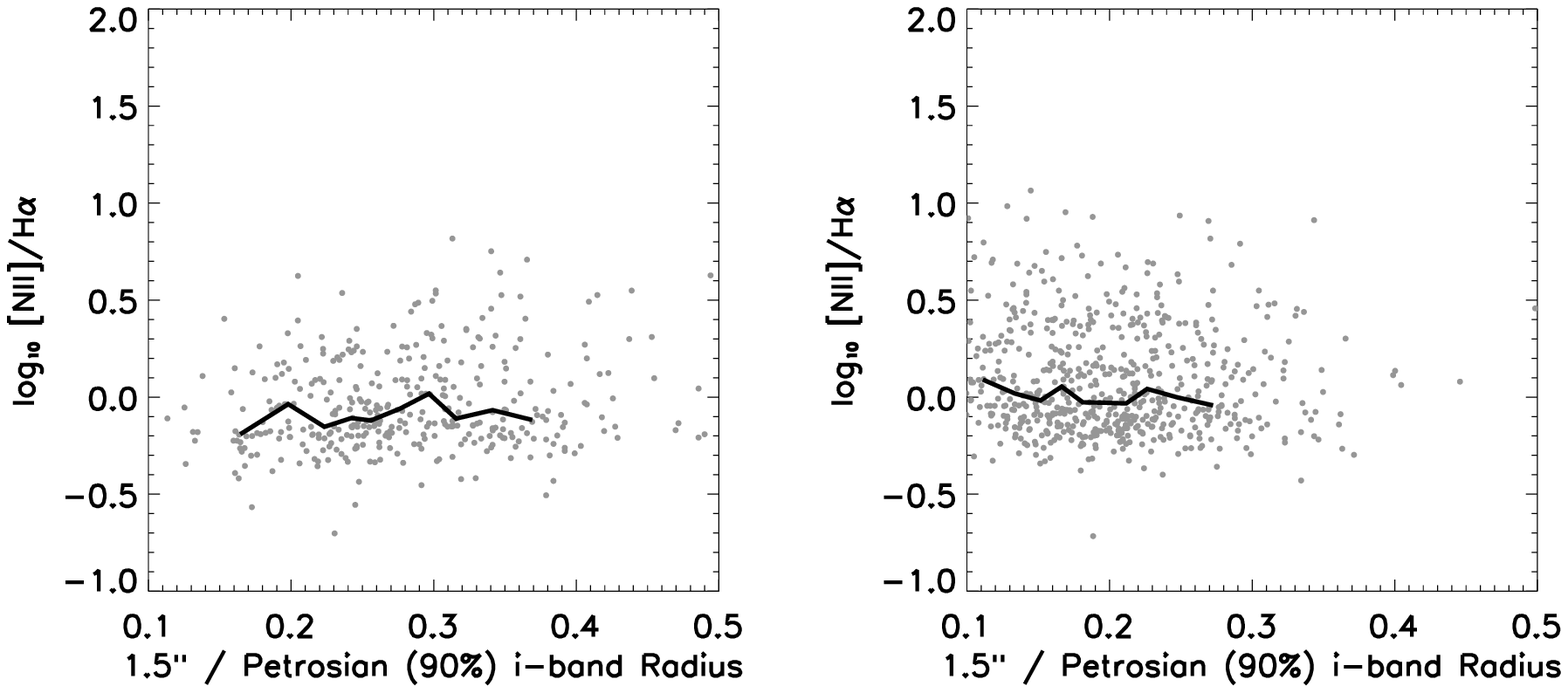}
\plotone{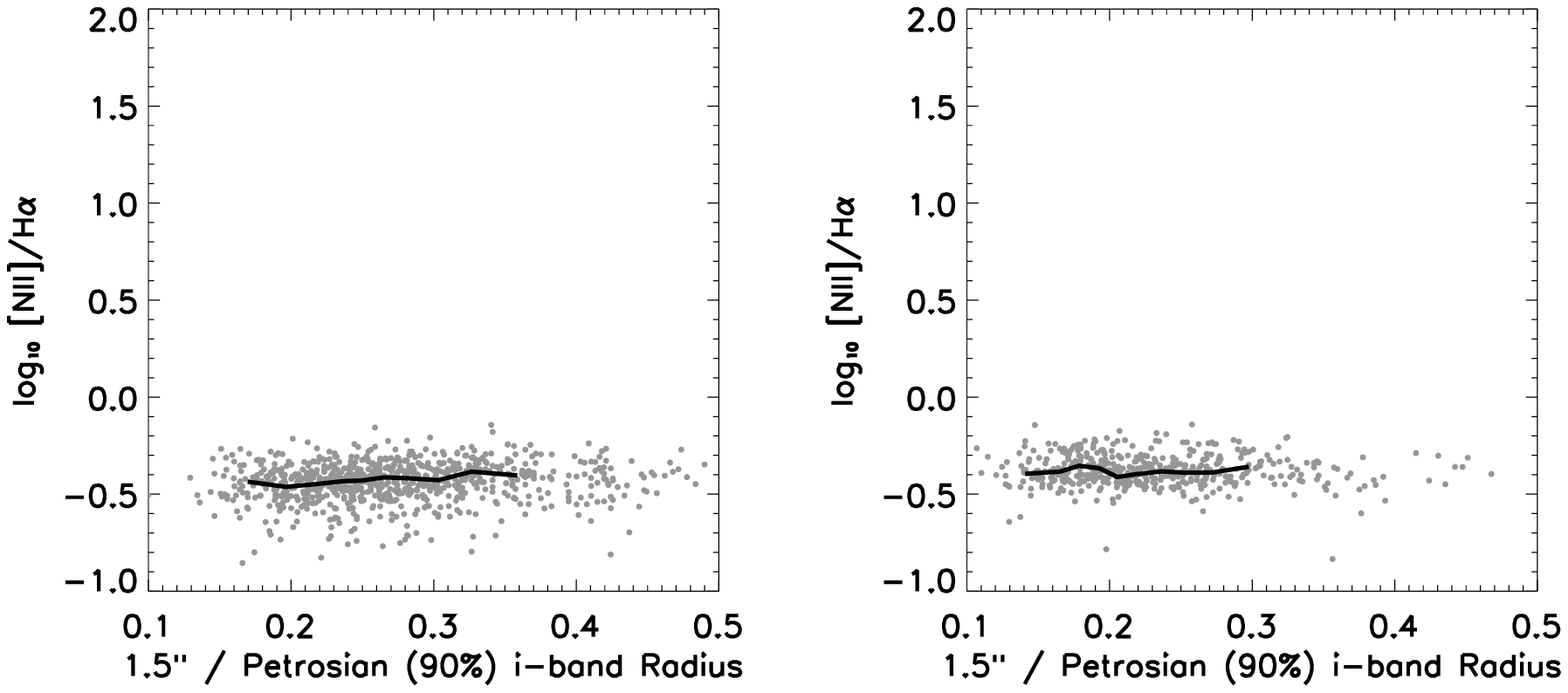}
\caption[]{(Top) The observed [NII]/H$\alpha$ line ratio, as a function of the
fraction of the total light within the SDSS fibre radius, for galaxies
classified as AGNs. The left--hand panel is for dim AGNs, while
right--hand panel is for AGN brighter than $M(r)$ = -20.75.  The solid line is
the median in each bin (with equal numbers of galaxies per bin).  (Bottom)
Likewise, The observed [NII]/H$\alpha$ line ratio, as a function of the
fraction of the total light within the SDSS fibre radius, for galaxies
classified as star--forming. The left--hand panel is for dim
star--forming galaxies, while right--hand panel is for star--forming galaxies
brighter than $M(r)$ = -20.75.  The solid line is the median in each bin (with
equal numbers of galaxies per bin).  }
\label{fig::han2_agn_v_radius}
\end{figure}

\begin{figure}
\plotone{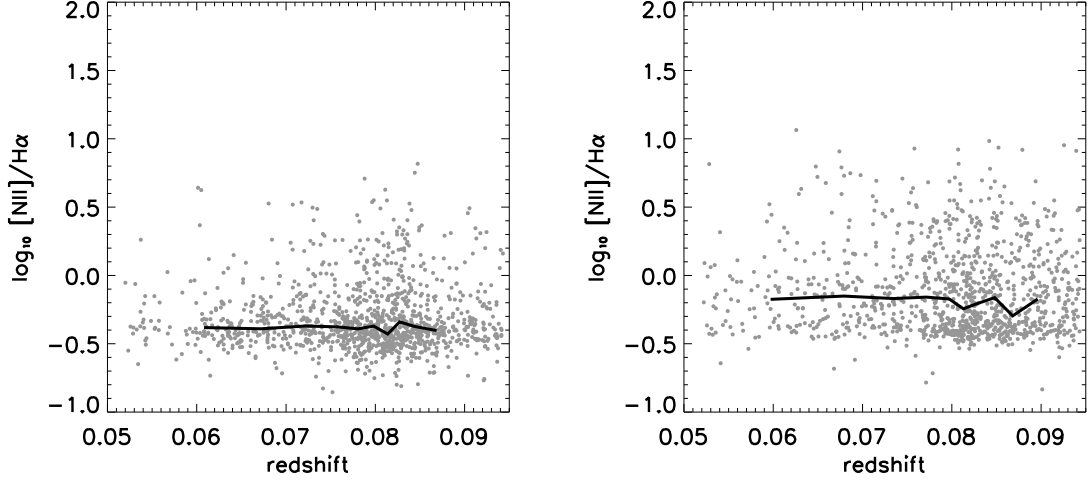}
\caption[]{The same as Figure \ref{fig::han2_agn_v_radius}, {\it i.e.}, the
measured [NII]/H$\alpha$ line ratio as a function of redshift for our galaxy
sample. (Left) Galaxies less luminous than $M(r)$ = -20.75. (Right) Galaxies
more luminous than $M(r)$ = -20.75. The solid line is the median in each bin
(with equal numbers of galaxies per bin)}
\label{fig::han2_vz}
\end{figure}

\begin{figure}
\plottwo{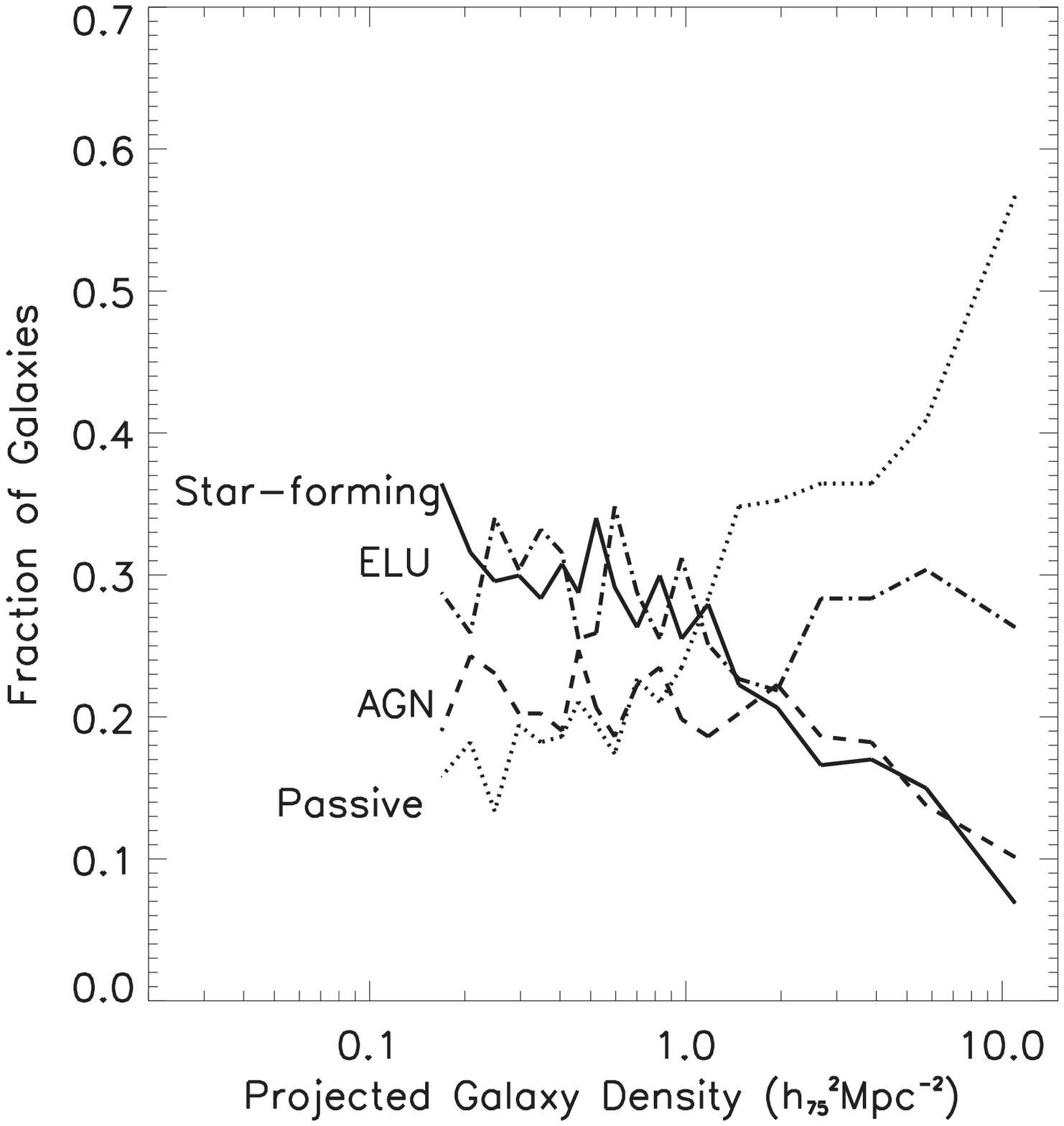}{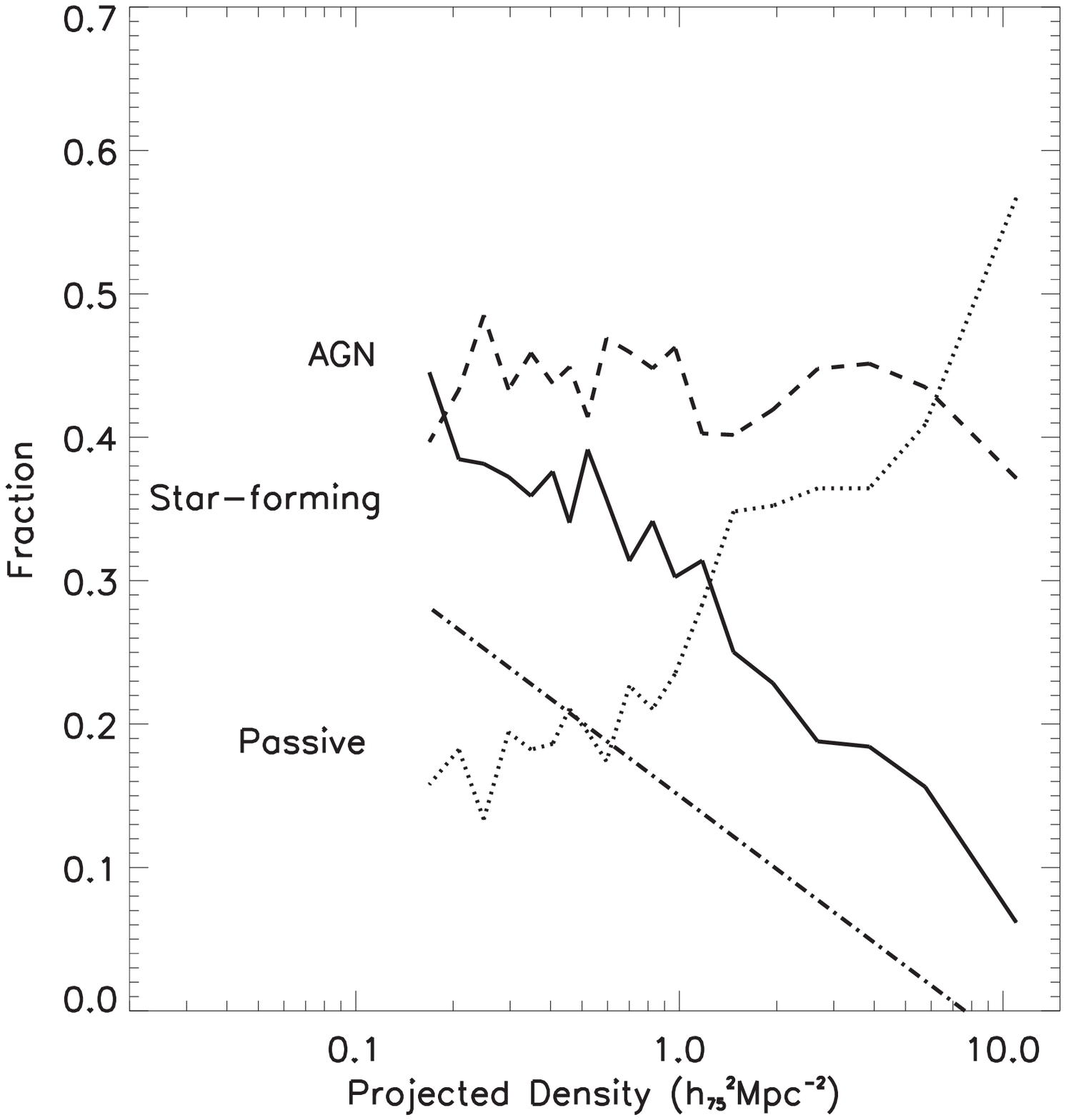}
\caption[]{(Left) The fraction of galaxies as a function of the different
classifications (Table 1, column 1), in our pseudo volume-limited sample, as a
function density. We see an increase in the fraction of passive galaxies with
density, and a decrease in the fraction of star--forming galaxies with
density. This is the SFR-density (Gomez et al. 2003) and morphology-density
(Dressler 1980; Dressler et al. 1997) relations. The fraction of galaxies
possessing an AGN is statistically--consistent with a constant over all local
galaxy densities probed here. (Right) We model the AGN/star--forming
distribution of ELU galaxies. This model (the dotted line) requires that the
overall fraction of star--forming galaxies in dense regions remains small. The
resultant AGN and star--forming fractions in this panel include galaxies
classified using the upper-limit technique and also the modeled ELUs. The
passive fraction remains unchanged.  }
\label{fig::agnfrc_dens}
\end{figure}

\begin{figure}
\plottwo{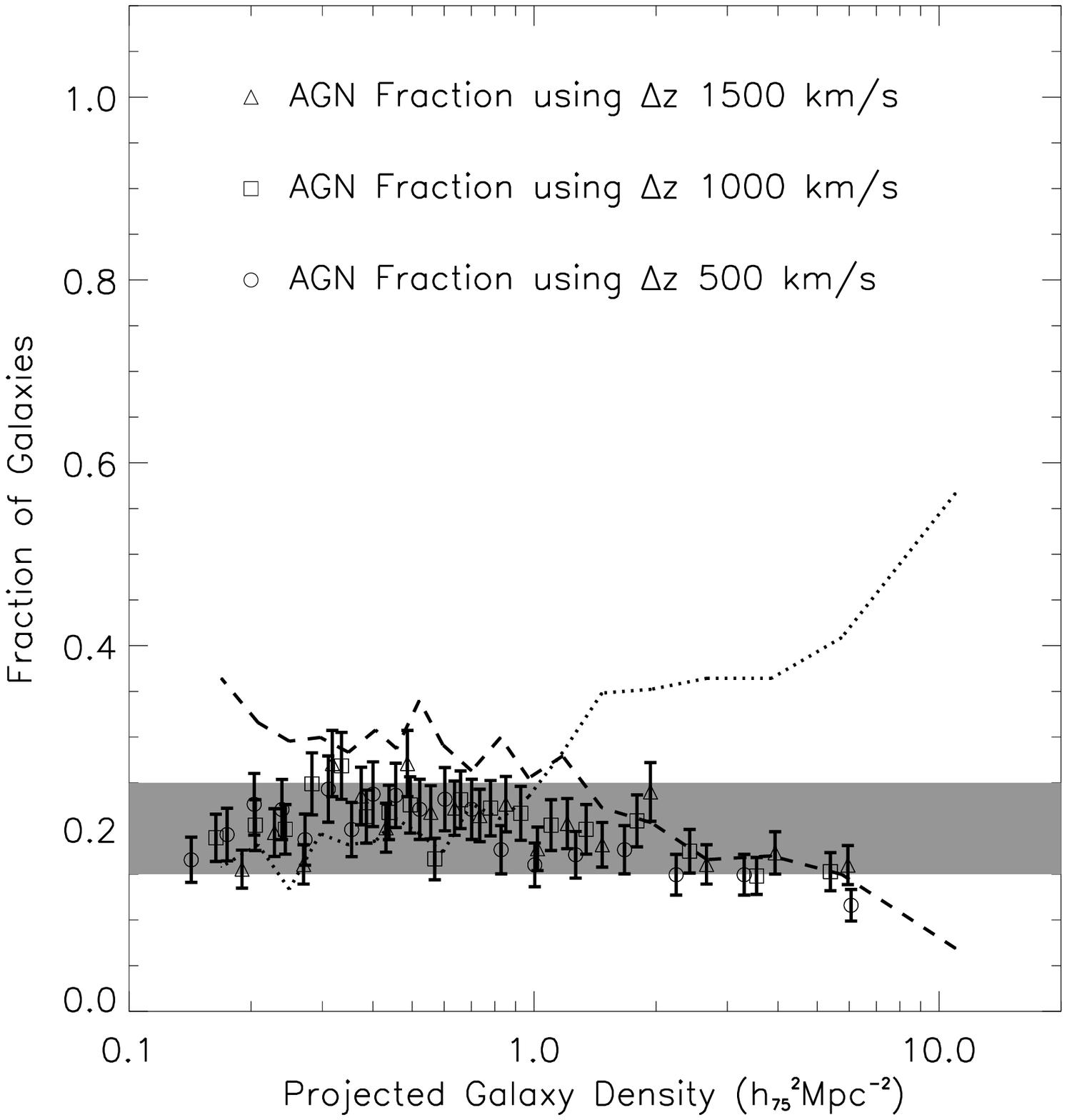}{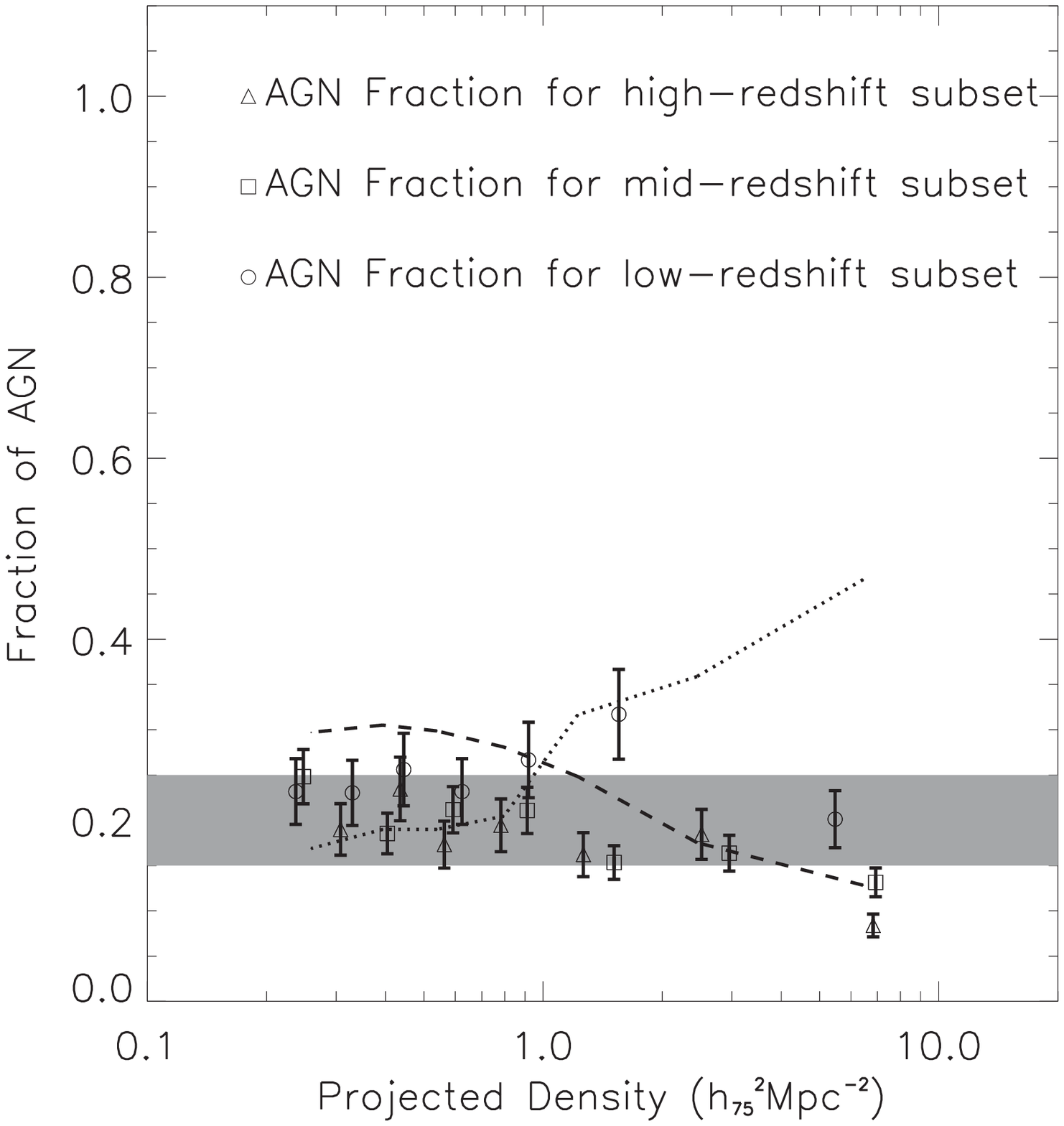}
\caption[]{(Left) We show the AGN fraction versus local galaxy density using
three different redshift shells within which to measure the projected
density. The open circles are using $\Delta z$ = 500 km/s, the squares are for
$\Delta z$ = 1000 km/s, and the triangles are for $\Delta z = 1500$
km/s. (Right) The AGN fraction in three near-equal volume redshift bins. The
circles are for $0.05\ge z < 0.078$, the triangles are for $0.078\ge z< 0.083$
and the squares are for $0.083\ge z< 0.095$ (see Gomez et al. 2003). The error
bars in this figure are 1/$\sqrt{N}$, where N is the number of galaxies in
each bin. The right--hand panel has approximately one--third of the total
galaxy sample in each redshift subset, and so there are fewer bins than in the
left--hand panel.  The dotted-line is the fraction of passive galaxies, while
the dashed-line is the fraction of star-forming galaxies.  The shaded region
encloses an AGN fraction of between 15\% and 25\% and thus covers the range of
AGN fractions discussed in this paper. A straight line (with zero slope) is a
good fit to the AGN Fraction data.}
\label{fig::bias}
\end{figure}

By using a combination of upper limits and detected emission lines, we compute
the line ratios for these ELU galaxies and place them on the usual diagnostic
diagram (Figure \ref{fig::diagnostic}).  If the ELU galaxies lie below the
model in Figure \ref{fig::diagnostic}, then it is reasonable to assume that
these galaxies are star--forming, as the true values for the missing \nii~
and/or \oiii~ emission lines must be smaller than their upper limits, which
would move the galaxies further into the star--forming region of Figure
\ref{fig::diagnostic}. We have used this technique to classify a further 230
of 1651 ELU galaxies as star--forming, bringing the fraction of
star--formation galaxies in our sample to 26\% (up from 21\% in Table 1). All
of these additional 230 galaxies were not originally classified using the
4--line method because they were missing a detection of the \oiii~ line (which
is a good indicator of AGN activity). These 230 star-forming galaxies have a
significantly smaller median H($\alpha$) EW of $\simeq 14$, compared to a
median H($\alpha$) EW of $\simeq 26$ for the 4--line classified star-forming
galaxies, {\it i.e.}, they have on average lower SFRs.

Likewise, we can classify ELU galaxies using the upper limits on any missing
\ha~ and/or \hb emission lines, and detections of the \nii~ and \oiii~
emission lines. In this case, if the observed line ratios for such ELU
galaxies place them above the model in Figure \ref{fig::diagnostic}, then it
is reasonable to assume that these ELU galaxies contain an AGN, as the true
value of \ha~ and/or \hb~ is smaller than the upper limit thus forcing the
galaxy further into the AGN region of Figure \ref{fig::diagnostic}. Using this
methodology we classify an additional 40 ELU galaxies as AGNs, bringing the AGN
fraction up slightly to 20\%. All of these galaxies were previously missed
because of weak \hb~ emission, which is a good indicator of recent
star--formation.

Using these new techniques, based upon the upper limits of missing
lines, we have reduced the fraction of ELU galaxies from 34\% to 28\%
of all galaxies in our sample. In the next section, we statistically
model the classifications of the remaining ELU galaxies.

\subsubsection{Statistical Modeling of ELU Galaxies}

Figure \ref{fig::agnfrc_dens} shows that $\sim 30\%$ of galaxies in the
densest environments are ELU galaxies. These ELU galaxies are either an AGN or
star--forming. It is now well established using complete galaxy samples
that the median SFR of galaxies in
the cores of clusters is low, and that the fraction of star--forming galaxies
is also low (see Gomez et al. 2003; CFGKM). Therefore, it is likely that most
of the ELU galaxies in dense environments are AGNs, and not
star--forming. Under this assumption, we can statistical model these ELU
galaxies and assign most of them in dense regions to our AGN fraction. In
creating this model, we first measure the fraction of unambiguous
star--forming galaxies over the total number of unambiguous star--forming
galaxies plus AGNs, versus density. Since the AGN fraction does not vary with
density, the slope of this fraction versus density is identical to the slope
of the star--forming fraction versus density (shown in Figure
\ref{fig::agnfrc_dens}). We then fix the amplitude of this relation such that
the fraction of star--forming galaxies in the densest regions is $5\%$, which
is similar to the fraction seen by CFGKM (in their highest density bin). Our
final model, {\it i.e.}, the fraction of ELU galaxies which are star--forming
versus density, is shown as the dash-dot line in Figure
\ref{fig::agnfrc_dens}.  We then add this modeled fraction of star--forming
and AGN ELU galaxies, as a function of density, to the existing fractions of
unambiguously classified star-forming galaxies and AGNs (from Section
\ref{identify} and Figure \ref{fig::agnfrc_dens}).  In Figure
\ref{fig::agnfrc_dens}, we show the final fractions of galaxies which are
classified as star--forming, AGN, and passive versus density for our total
sample of 4921 galaxies. We note that 82\% of the ELU galaxies, which were not
classified using the upper limits on the emission lines, were statistically
classified as AGNs, while the remainder are star--forming.

\subsubsection{Testing our Model}
\label{testingI}

\begin{figure}
\plottwo{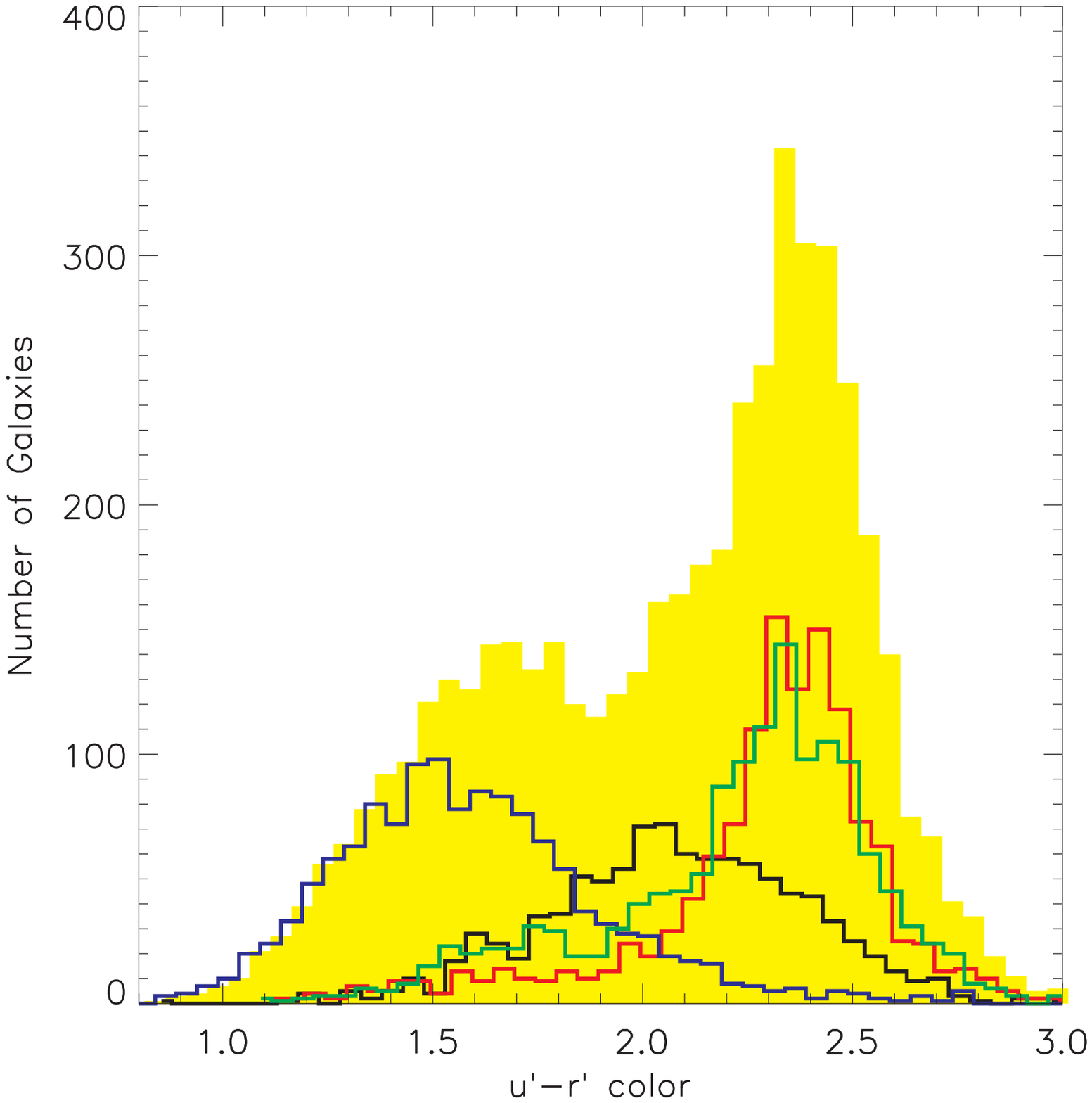}{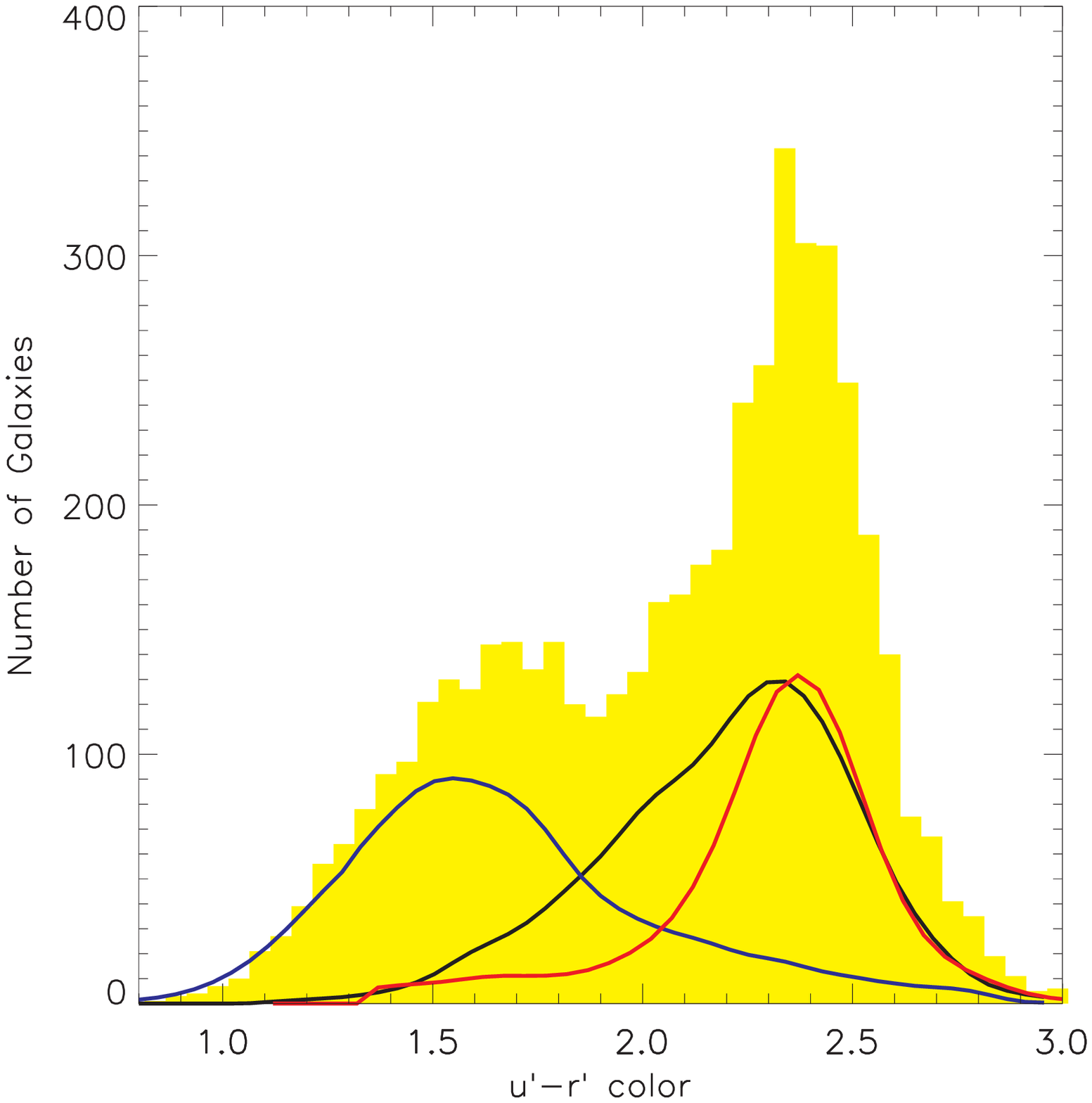}
\caption[]{(Left) The distribution of $u^*-r^*$ color for the different
classifications of galaxies used herein. The yellow area is for all
4921 galaxies in our sample. The red line is for passive galaxies. The
blue line is for star--forming galaxies. The black line is for
unambiguously identified AGN (4--line method), while the green line is
for the ELU galaxies.  As expected, most passive galaxies are red and
most star--forming galaxies are blue. (Right) We now model distribution
of color for the AGN and star--forming color distribution of ELUs as
in Figure \ref{fig::agn_color_model}.  The AGN and star--forming
populations now include these ELUs.  }
\label{fig::agn_color_hist}
\end{figure}

Before we interpret our results, it is fair to question the reliability of our
empirical modeling.  We test our model using the SDSS photometric imaging data
via the colors of our galaxies. In particular, we use the $u^*- r^*$ rest-frame color as an
indicator of on--going star--formation, as the $u$-band flux from these
galaxies will be dominated by young, massive stars (see Strateva et
al. 2001). The colors also have the advantage of measuring the total spectral
energy distribution of the whole galaxy (we use Petrosian magnitudes here) and
are therefore, unaffected by fiber aperture effects and stellar absorption
problems. In Figure \ref{fig::agn_color_hist}, we show the distributions of
color (SDSS $u^*- r^*$) for all galaxies in our sample. The star--forming and
passive galaxy populations follow the expected trends, {\it i.e.}, the
star--forming galaxies are the bluest (in $u^*- r^*$) galaxies in our sample,
while the passive galaxies are the reddest. The AGN population lie between
these two populations; redder than the star--forming galaxies, but slightly
bluer than the passive galaxies.


Figure \ref{fig::agn_color_hist} shows the distribution of color for the ELU
galaxies which could not be identified using upper limits on the missing
emission line, {\it i.e.}, the sample of ELU galaxies that we statistically
classified. Clearly, the bluer ELU galaxies are likely to be star--forming,
while the redder ELU galaxies are likely AGNs as, by definition, their spectra
contains at least one significant emission line. As a test of our
classification methods, we note that 90\% of the 230 ELU galaxies we
classified as star--forming using the upper limits in Section 3.1.1 have $u'-
r^*$ colors of $<2.0$, which is fully consistent with the observed colors of
our unambiguously classified star-forming galaxies. This is strong
confirmation that these ELU galaxies are truly star--forming and the methods
we employ are robust.

\begin{figure}
\plotone{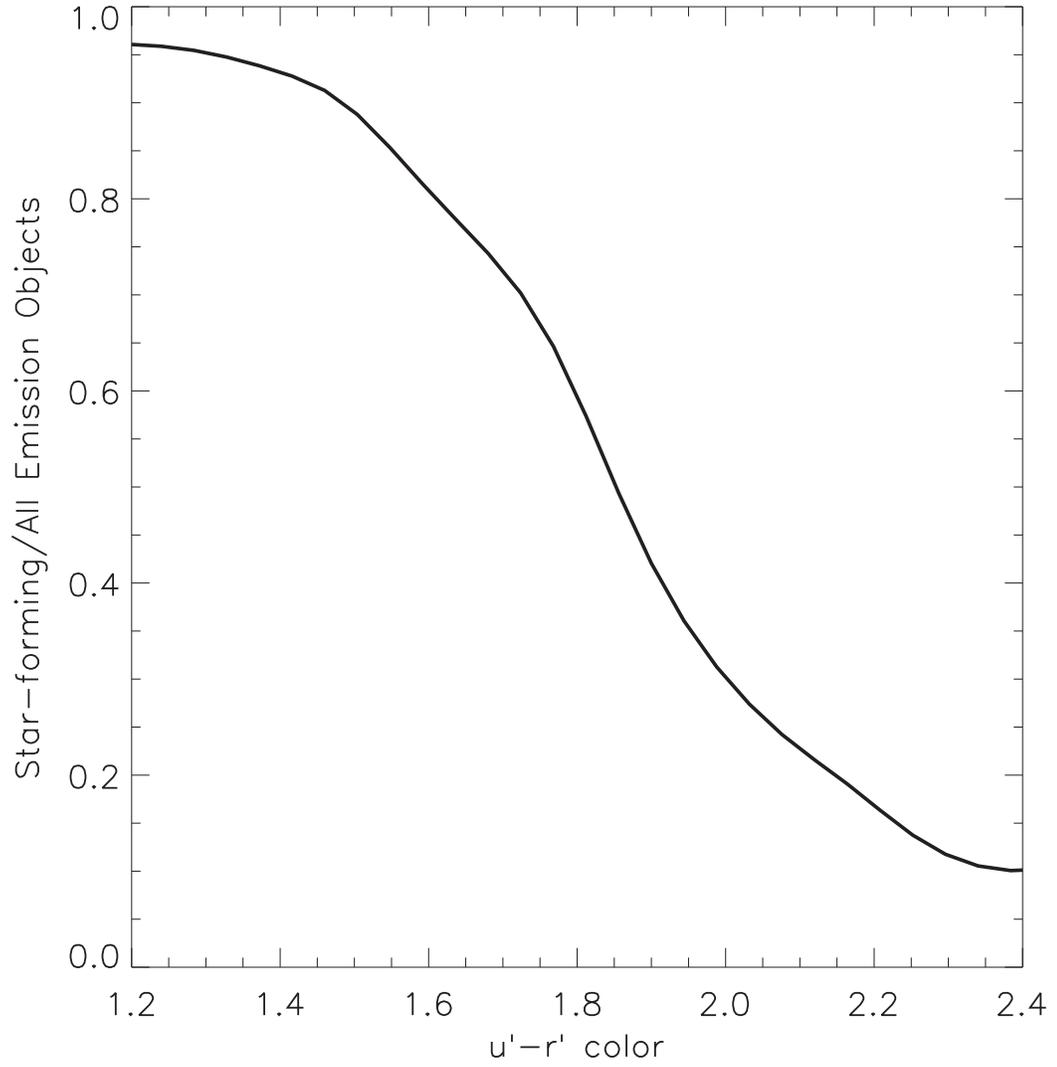}
\caption[]{The fraction of unambiguously classified star--forming
galaxies over the total number of unambiguously classified galaxies in
the sample, as a function of their observed $u^*-r^*$ color. This
demonstrates that most of the blue emission line galaxies are
star--forming galaxies, while the red emission line objects are
AGN. We use this observed fraction to statistically classify the
remaining ELU galaxies.  }
\label{fig::agn_color_model}
\end{figure}

We can also model the distribution of star--forming galaxies and AGNs in the
ELU galaxies by assuming that the color distribution for the fraction of
unambiguous star--forming galaxies (4--line method) over all unambiguously
classified galaxies, is correctly measured in our sample. This fraction is
shown in Figure \ref{fig::agn_color_model} and simply states that nearly all
galaxies with $u^*-r^* < 1.5$ are star--forming galaxies, while above $u^*-r^* >
2$, nearly all our unambiguously classified galaxies possess an AGN. We apply
this model to the sample and show in Figure \ref{fig::agn_color_hist} (right)
the resulting color distributions for our classifications.  This model
identifies 77\% of the ELU galaxies as AGNs, while the remaining 23\% are
identified as star--forming. These observed fractions, which are based solely
on the total colors of these ELU galaxies, are consistent with our model
described above.


Finally, we considered here the effects of dust extinction (reddening) on our
observed color distributions. For example, dusty star--forming galaxies could
mimic the colors of our unambiguously classified AGNs and would cause an
over--estimation of the AGN fraction in the ELUs galaxies. To quantify this
problem, we use the Bruzual \& Charlot (2001) spectral synthesis models to
estimate the amount of dust extinction and metalicity required to redden the
$u^*-r^*$ color of a star--forming galaxies such that it would be
mis--classified by our methods as an AGN. We begin by calculating the $u^*-r^*$
color for a $z=0.07$ star--burst galaxy 0.3 Gigayears after its burst and with
a dust extinction similar to that of our Galaxy (see Charlot \& Longhetti 2001
for these parameters). The predicted $u^*-r^*$ color of such a galaxy is fully
consistent with the observed $u^*-r^*$ colors of our star--forming galaxies
($u^*-r^*=1.5$).  We then increased the dust extinction, keeping all other
parameters fixed, until we obtained the mean $u^*-r^*$ color observed for our
AGNs. We find that we require an additional 4 magnitudes of dust extinction
(beyond that seen in our Galaxy) to result in a mis--classification of these
star--forming ELU galaxies. This is a large amount of extinction and is
unlikely to affect 25\% of bright galaxies in the local Universe which are
under--going active star--formation. We therefore conclude that it is highly
unlikely that star--forming galaxies make up a large portion of the ELU
galaxies.

\subsection{AGN Fraction as a Function of Galaxy Morphology}
\label{morphology}

\begin{figure}
\plotone{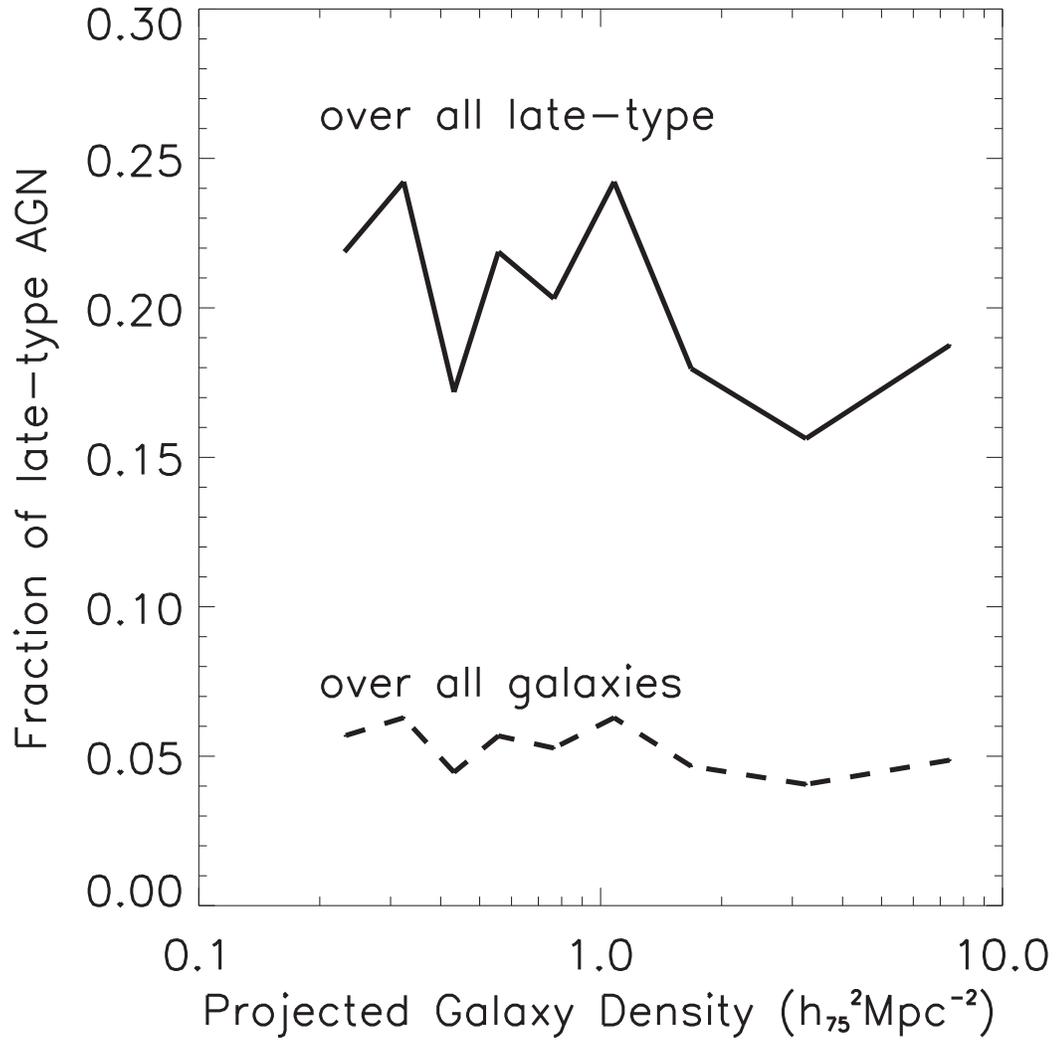}
\caption[]{The blue line shows the fraction of just late--type
(spiral) galaxies that have an AGN. While the red line shows the
fraction of all galaxies that are both late--type and have an AGN.
This figures shows that late--type galaxies have the same AGN fraction
as all galaxies, and there is again no dependence on density.  }
\label{fig::agnfrc_dens2_i}
\end{figure}

In this section, we study the fraction of AGNs in galaxies as a function of
the morphology of the host galaxy. First, we utilize the inverse concentration
index ($C_{in}$) in the SDSS $r$ band, which is defined to be
$r_{50}/r_{90}$, where $r_{50}$ is Petrosian radius containing 50\% of the
light and $r_{90}$ is Petrosian radius containing 90\% of the light (see
Stoughton et al. 2002; Strateva et al. 2001). Therefore, a galaxy with a
diffuse light distribution (a spiral galaxy) would have a high concentration
index, while compact galaxies have a low index value. As in Gomez et
al. (2003), we use a threshold $C_{in} \ge 0.4$ to identify late--type
(spiral) galaxies, which is higher than the $C_{in} \ge 0.33$ threshold
proposed by Shimasaku et al. (2001) and Strateva et al. (2001). We prefer the
higher threshold in $C_{in}$ as it provides a purer, yet more incomplete,
sample of spiral galaxies than those advocated by Shimasaku et al. (2001) and
Strateva et al. (2001). In Figure \ref{fig::agnfrc_dens2_i}, we present the
AGN fraction for galaxies with $C_{in}\ge0.4$ (late--type galaxies) as a
function of local galaxy density. As seen in Figure \ref{fig::agnfrc_dens2_i},
we observe no trend with density for just the late--type (spiral) galaxies and
the fraction of all late--type galaxies with an AGN is 20\%.


To study the fraction of early--type galaxies with an AGN, we can not use
$C_{in} < 0.4$ as this sample is severely contaminated by late--type galaxies
(see Gomez et al. 2003; Shimasaku et al. 2001; Strateva et al. 2001). As an
alternative, we can study the subset of galaxies that reside in known clusters
of galaxies, as these galaxies will be predominately early--type (ellipticals
and S0's) galaxies. To achieve this goal, we use the C4 cluster catalog which
was described in Gomez et al. (2003) and Nichol (2003).  In this paper, we use
a sample of 23 C4 clusters in the EDR area that have a mean redshift between
$0.053 \le z \le 0.092$.  We then select 967 galaxies from our original sample
that lie within a projected distance of 2 virial radii from one of these
clusters, and within $|z_{galaxy} - z_{cluster}| \le 4 \sigma_v$, where
$\sigma_v$ is the velocity dispersion of the clusters. In Table 1, we provide
the fraction of galaxy as a function of our classifications (passive,
star--forming and AGN) for these 967 cluster galaxies.  As expected, the
fraction of star--forming galaxies is lower for these cluster galaxies than
for our full sample (see Table 1). As seen in Gomez et al. (2003), the
fraction of star--forming galaxies becomes significantly lower in the inner
(projected) 500kpc of these 23 clusters. Likewise, the fraction of passive
galaxies (no detected emission lines) is higher in clusters than for the full
sample, and continues to increase towards the inner (projected) 500kpc of our
clusters. However, we again observe that the fraction of galaxies with an AGN
remains nearly the same in our 23 clusters as for our full sample, and we only
witness a small decrease (which is statistically insignificant) in the
fraction of AGNs within the inner (projected) 500kpc of our clusters.

We note that we have
not modeled the ELU galaxies in our cluster analysis (as we did in Section \ref{density}
for the galaxy densities).
However, it is worth noting in Table 1 columns 2 and 3, that there is large number of
ELUs in the cores of clusters. As we expect most of these ELUs
to be AGNs, a model for cluster ELUs similar to the one
used in Section \ref{ELU} would have the same consequences as shown in Figure \ref{fig::agnfrc_dens}:
raising the total fraction of AGNs and keeping the fraction of AGNs constant.

\begin{deluxetable*}{cccc}
\tablewidth{0pt}
\tablecaption{\bf Galaxy Type Fractions}
\tablehead{
\colhead{Galaxy Type} & \colhead{Fraction (all galaxies)} & \colhead{Fraction $< 2$ R$_{virial}$} & \colhead{Fraction $< 0.5$ R$_{virial}$} \\
\colhead{} & \colhead{(cluster and field)} & \colhead{(cluster)} & \colhead{(cluster)}}
\startdata
AGN & 19\% & 17\% & 14\%  \\
Star--forming &21\% & 16\% & 9\% \\
Passive & 26\% & 35\% & 46\%\\
ELU &34\% & 32\%  & 32\% \\
\enddata
\end{deluxetable*}

\subsection{AGN Fraction Summary}
In Figure \ref{fig::agnfrc_dens}, we show the AGN fraction as a
function of local galaxy density for our pseudo volume--limited sample
of galaxies. In this figure, we present two versions of the {\it
AGN--density} relation. The left--hand plot represents the absolute
lower limit on the AGN fraction, as we only plot our unambiguously
classified AGNs and assumes all the unclassified emission line
galaxies are star--forming. This is clearly unreasonable. The
right--hand plot presents the AGN fraction as a function of local
galaxy density when we use our final AGN classifications based on all
the methods discussed herein, {\it i.e.}, both the 4--line and 2--line
method, as well as the upper limits and statistical modeling of the
ELU galaxies. This is a more realistic representation of the true {\it
AGN--density} relation. In both cases however, our {\it AGN--density}
relation is consistent with a constant AGN fraction over two orders of
magnitude in local galaxy density.

We have extensively tested our classifications of the ELU galaxies.
We find that the fraction of ELU galaxies which contain an AGN is
$\sim 75\%$, which is consistent with the model we
employed in Figure \ref{fig::agnfrc_dens}, which is based on the
environmental dependence of unambiguous star--forming galaxies (the
dotted line) . Therefore, our classifications are robust and
internally consistent, {\it i.e.}, the color, environment and emission
lines of these ELU galaxies all point to the same classification.  In
total, we find that $\simeq 40\%$ of galaxies in our
pseudo-volume-limited sample contain an AGN, which is remarkably
consistent with the 43\% measured by Ho et al. (1997).
We note that previous samples have different luminosity limits and
selection algorithms and
these must be accounted for in any detailed comparison of AGN fractions.
Also, we remind the reader that this value of 40\% is only valid
in the context of the model we apply (i.e. that most ELUs in clusters
are {\it not} star-forming according to the known star-formation density
and star-formation-fraction density relation

\section{Discussion}
\label{discussion}

\subsection{Comparisons to Previous Works}

Based on our sample of unambiguously identified AGN, the overall
fraction of galaxies with an AGN we observe (20\%) is similar to that
reported by CFGKM, who found that 17\% of their magnitude limited
sample of $\sim$3500 galaxies ($m_R \le 15.4$) possessed an AGN. We
note that the AGN classification process used by CFGKM is very similar
to ours, while the galaxy samples are nearly independent. Carter et al. (2001)
further suggested that as many as 28\% of their sample could be AGN,
if they classify all of their unclassified emission line galaxies as
AGN.  By modeling our unclassified emission line galaxies (Section
\ref{identify}), we find that $\sim$ 40\% of our sample could be
AGN. This larger fraction is closer to that found by Ho, Filipenkko \&
Sargent (1997) who used the Palomar survey of galaxies to identify
AGNs.

The fraction of galaxies in our sample that definitely harbor an AGN
($20\%$) is significantly higher than found in the earlier studies of
Dressler et al. (1984) and Huchra \& Burg (1992). We also find our
fraction to be significantly larger than that reported by Ivezic et
al. (2002), who found that 5\% of SDSS galaxies had an AGN. This could
be the result of the different AGN classification techniques, {\it
e.g.} Ivezic et al. (2002) required at least four $>3\sigma$ detected
emission lines for each galaxy, and used a higher threshold for
separating AGNs from star--forming galaxies than used here.
Interestingly, Ivezic et al. (2002) finds
that by only using those SDSS galaxies which have radio match
(ignoring their spectral AGN classifications), then the potential AGN
fraction rises to 30\%, more in agreement with our results.  All
together, these works suggest that at least $\sim 20\%$ of luminous
galaxies have an AGN, but the true fraction is probably closer to
$\sim 40\%$. We note that we have not performed a fully detailed comparison
of previously measured AGN fractions. Such an analysis would need to take
into account the variety of selection differences and the incompleteness
in their respective samples.

We find that the fraction of galaxies with an AGN is independent of
environment. This result is consistent with the works of Coziol et
al. (1998), Shimata et al. (2000), Monaco et al. (1994), and Schmitt
(2001), who use a variety of techniques to quantify environment of
AGNs.  Our results are also in agreement with the recent analysis by
CFGKM, who use a similar method for measuring the local galaxy
density. We also show that the AGN fraction in clusters is the same as
for all galaxies, which is different from Dressler et al. (1984) who
find five times as many AGN in the field as in clusters. We attribute
this difference to the selection criteria used by Dressler et al.,
specifically they missed both the \nii~ and $H\alpha$ emission lines
(see also Way, Flores, and Quintana 1998).

Our results are also consistent with the recent quasar clustering
analyzes of Croom et al. (2002) and Oltram et al. (2002) for the 2dF
QSO Redshift Survey (2QZ). These authors find that the clustering of
luminous quasars (at a mean redshift of $z=1.4$) is identical to that of local
galaxies, thus suggesting that these two populations (galaxies and
quasars) where formed about the same time and follow the same density
distribution. This conclusion is re--enforced by Cattaneo \& Bernardi
(2003) who use the stellar ages of SDSS ellipticals to constrain the
cosmic accretion history of supermassive black holes.  Taken together,
these works clearly indicate that AGNs (including quasars) could be
unbiased tracers of the galaxy distribution at both low and high
redshift. However, the Martini et al. (2002) study of X-ray 
classified AGN notes that many optical AGN in cluster may be obscured by
dust. If this were the case, one could not recover 100\% of the AGN
population using optical studies alone. 

\subsection{Physical Interpretation}

In Figure \ref{fig::agnfrc_dens}, we confirm the findings of Gomez et
al. (2003) and Lewis et al. (2002) that the SFR--density relation has
a ``critical density'' at $\simeq 1 h^{-2} Mpc^{2}$, {\it i.e.} below
this density the fraction of galaxies that are star--forming is high
and constant (see Postman \& Geller 1984), while above this density
the fraction drops rapidly. Likewise, we see the opposite trend for
the passive galaxies (with no emission lines) in that the fraction of
such galaxies increases rapidly above the critical density of $\simeq
1 h^{-2} Mpc^{2}$, {\it i.e.}, consistent the {\it
morphology--density} (Dressler et al. 1997; Postman \& Geller
1984). It is therefore surprising that the AGN population has no
apparent {\it AGN--density} relation, as other properties of galaxies
(morphology, luminosity, star--formation rate {\it etc.}) show a
strong dependence on local galaxy density (Gomez et al. 2003; Hogg et
al. 2002). We also find no {\it AGN--density} relation for the early--
and late--type galaxies.

These observations can be naturally understood under the hypothesis
that the AGN population is primarily tracing only the bulge component
of galaxies.  This hypothesis would explain why we see no dependence
on local galaxy density, as most galaxies have a bulge, and why there
is no correlation with morphological type, as the disk component is
irrelevant to the existence of an AGN. Such a model would then suggest
that star--formation in the disk of a galaxy is not driven by the
presence of an AGN, and vica versa. This hypothesis is also consistent
with the fact that most of our AGNs appear to be redder in their
overall $u^*-r^*$ color) than known star--forming galaxies.
If there were a strong
relationship between the existence of an AGN and the ability of a galaxy to
form stars, we might expect the AGN fraction to decrease with density, as the
star-forming fraction is seen to descrease with density.

Our hypothesis is supported by the observations of Ho et al. (1997)
and Martini et al. (2002) who see a strong correlation between the
properties of AGNs and the bulge of their host galaxies. Furthermore.
Kormendy \& Gebhardt (2001) observe that the mass of the central
supermassive black hole in nearby galaxies is independent of the
luminosity of the disk component of a galaxy and only correlates with
the properties of the galactic bulge.  This hypothesis is also
consistent with the idea that AGNs (including quasars) were formed at
the same epoch as the old stellar populations in the bulges of these
galaxies (Cattaneo \& Bernardi 2003).

We should note however, that sub--populations of AGNs are known
to be strongly correlated with the morphology of the host galaxy; for
example, Seyfert galaxies preferentially reside in spirals and S0s,
while radio--loud galaxies are predominantly elliptical galaxies (see
Ivezic et al. 2002). Therefore, these sub--populations of AGNs would
follow the same density relations as their host galaxies, {\it i.e.},
the fraction of strongly star-forming galaxies in the cores of
clusters is much lower than that seen in the field population. We will
investigate such sub--populations of AGNs in our future work; however
it is intriguing to note that when the classifications of these
sub--populations is ignored (and we simply classify galaxies as AGN or
star--forming) the total AGN population becomes independent of local
galaxy density. This indicates that the fraction of galaxies
classified in these different sub--populations (Seyferts, LINERs {\it
etc.}) appear to fortuitously sum to a constant over two orders of
magnitude in local density.

Our work strongly supports the hypothesis that the bulge of every
luminous galaxy, regardless of its overall morphology and environment,
contains a supermassive black hole (BH) which is fueling the AGN (see
the discussion by Kormendy \& Gebhardt 2001). In this model, the black
holes, and their mass function, is established at an early
cosmological epoch, {\it e.g.}, through early formation of the
galactic bulges in these galaxies (via merger events) and each of
these bulges was seeded by BH. This
model was recently investigated by Di Matteo et al. (2003) using
detailed hydrodynamical simulations to trace the growth and activity
of supermassive black holes in galaxies. They found that a majority of
the BH mass is accreted by $z\sim3$ and no further growth happens at
later epochs, similar to proposed evolution of galactic bulges. They
also found that the BH accretion rate (BHAR) density evolution follows
the SFR density evolution to z = 4, but then drops rapidly after that
epoch resulting in the BHAR being 100 times lower than the SFR density
by $z=1$. Therefore, the BHAR plays no role in the evolution of
galaxies at low redshift. In the near future, we should be able to
test such predictions using our data from the SDSS.

Under the assumption that every luminous galaxy in our sample contains
a supermassive BH, the fraction of galaxies we observe with an AGN
($\simeq40\%$) must therefore be dependent upon the lifetime (or
``duty cycle'') of AGNs (or is dependent on the orientation of the
torus surrounding the BH). For example, the interval of look--back
time spanned by the redshift range of our sample is $5.7\times 10^8$
years (from $z=0.05$ to $z=0.095$). We note that our AGN fraction
shows no sign of changing with redshift over this small redshift
interval. Therefore, to explain a constant observed AGN fraction of
40\%, the average lifetime of these AGNs could be as high as $0.4
\times 5.7\times 10^8 \simeq 2\times 10^8$ years. This is longer than
previous estimates of the lifetime of an AGN, {\it e.g.}, a few
million years (see Martini et al. 2002), and is longer than the
Salpeter lifetime of $4\times 10^7$ years (Salpeter 1964). However, it
is more consistent with the estimates on the lifetime of quasars from
quasar clustering measurements (see Haiman \& Hui 2001). One
explanation for this longer lifetime is that most of our AGN
detections are LLAGNs, where the accretion rate onto the BH is lower
than for a normal AGN, thus prolonging the lifetime of the AGN
activity.

Alternatively, our AGNs may be experiencing several bursts of activity
over the interval of look--back time probed by our sample, {\it i.e.},
$0.4 \times 5.7\times 10^8 \sim n\,\times\, \tau$, where $\tau$ is the
lifetime of the AGN and $n$ are the number of bursts. If we assume
$\tau\simeq 10^7$ years in agreement with previous observations and
the Salpeter time, then this implies $n\sim40$ bursts between $z=0.05$
and $z=0.095$.  In fact, these arguments are simply a re-statement of
the fact that we see a high fraction of galaxies with an AGN, so they
must be a common phenomenon either because they live a long time, or
burst many times. In either case, our observed high fraction of
galaxies with an AGN appears inconsistent with a merger--driven model
for the fueling of the AGN (Gunn 1979; Kauffmann \& Haehnelt 2002), as
this would imply a high merger rate for local galaxies. We also see no
dependence on environment, which would be expected if AGN activity was
primarily driven by galaxy mergers.

\section{Conclusions}
\label{conclusions}

In this paper, we have studied the fraction of galaxies that possess
an AGN as a function of both environment and galaxy type. We have used
the Early Data Release of the Sloan Digital Sky Survey to define a
pseudo volume--limited sample which is brighter than $M^{\star} +
1.45$, within a redshift range of $0.05 \le z \le 0.95$. Using these
data we conclude:

\begin{itemize}
\item{The overall fraction of galaxies with an AGN is at least $20\%$
(with an unambiguous classification), but maybe as high as $40\%$ if
we have correctly modeled the unclassified emission line galaxies in
our sample. This confirms the earlier observations by
CFGKM and Ho et al. (1997);}

\item{Our results are robust against measurement error on the emission
lines, stellar absorption and technique to classify galaxies;}

\item{Over two orders of magnitude, the fraction of galaxies with an
AGN is independent of the (projected) local galaxy density (see
CFGKM). This is in contrast to star--forming galaxies which are seen
to decrease with density (Gomez et al. 2003) and passive galaxies that
increase with density (Dressler et al. 1997). Therefore, AGNs are
unbiased tracers of the overall galaxy population;}

\item{The AGN fraction is also independent of the overall
morphological type of the host galaxy, and therefore, there appears to
be no overall relationship between the star--formation activity in the
disk component of galaxies and the presence of an AGN. A plausible
interpretation of this result is that the AGN phenomenon is related to
the bulge component of galaxies, which is consistent with the
hypothesis that a supermassive black hole resides in the bulge of all
galaxies (see Kormendy \& Gebhardt 2001);}

\item{The high fraction of galaxies with an AGN suggests that the
lifetime (or ``duty--cycle'') of these AGNs is long, or that they burst
often.  Using the interval in look--back time spanned by our sample, we
estimate that the lifetime of AGNs is $\simeq 2\times 10^8$ years,
which is significantly longer than the Salpeter time. Alternatively,
if the lifetime of the AGNs is $10^7$ years, then these AGN burst on
average $\sim40$ times between $z=0.095$ and $z=0.05$. If the AGN are
merger--driven, then this implies a very high merger rate which is
inconsistent with other observations and models.}
\end{itemize}


The authors would like to thank Tiziana Di Matteo, Kathy Romer,
Ruper Croft, and Tomo Goto for their help and insightful discussions.  AMH
acknowledges support provided by the National Aeronautics and Space
Administration (NASA) through Hubble Fellowship grant
HST-HF-01140.01-A awarded by the Space Telescope Science Institute
(STScI).

Funding for the creation and distribution of the SDSS Archive has been
provided by the Alfred P. Sloan Foundation, the Participating
Institutions, NASA, the National Science Foundation, the
U.S. Department of Energy, the Japanese Monbukagakusho, and the Max
Planck Society.  The SDSS Web site is http://www.sdss.org/.

The SDSS is managed by the Astrophysical Research Consortium (ARC) for
the Participating Institutions. The Participating Institutions are the
University of Chicago, Fermilab, the Institute for Advanced Study, the
Japan Participation Group, The Johns Hopkins University, Los Alamos
National Laboratory, the Max-Planck-Institute for Astronomy (MPIA),
the Max-Planck-Institute for Astrophysics (MPA), New Mexico State
University, the University of Pittsburgh, Princeton University, the
United States Naval Observatory, and the University of Washington.

\end{document}